\documentclass[a4paper, 10pt, aps, prb, twocolumn, showkeys, superscriptaddress]{revtex4-1}

\usepackage[nomath]{stix}
\usepackage{newtxmath}	
\usepackage[T1]{fontenc}
\usepackage[utf8]{inputenc}			
\usepackage{mathtools, amsfonts}	% The 'mathtools' package automatically loads 'amsmath'. Similarly 'amssymb' also loads 'amsfonts'
\usepackage[activate={true, nocompatibility}, factor=1100, kerning=true]{microtype} 
\usepackage{graphicx}
\usepackage{siunitx}
\usepackage[tiny,center,compact]{titlesec}
\usepackage[a4paper, portrait, margin=0.7in]{geometry}
\usepackage{pdfpages}
\usepackage{pgffor}
\usepackage{pax}
\usepackage[bookmarks=false]{hyperref} 	
\hypersetup{
colorlinks=true,
linkcolor=black,
citecolor=blue,
urlcolor=blue,
pdfstartview={FitH},		
pdfauthor={Achint Jain},
pdftitle={One-dimensional edge contacts to a monolayer semiconductor}
}

\titlespacing{\section}{0pt}{1.5ex}{1.2ex}					% To define the spacing between a section header and its contents 
\makeatletter
\AtBeginDocument{\let\LS@rot\@undefined}					
\makeatother

\begin{document}

\def \mos2{MoS$_{2}$}
\def \mose2{MoSe$_{2}$}
\def \mote2{MoTe$_{2}$}
\def \ws2{WS$_{2}$}
\def \wse2{WSe$_{2}$}
\def \sio2{SiO$_{2}$}
\def \uvo3{UV-O$_{3}$}
\def \moo3{MoO$_{\mathrm{x}}$}
\def \Id0{$I_{_\mathrm{D}}$}	% Here double subscript is used to make the capital letters D, G and T in V_D, V_G, V_T and I_D appear smaller.
\def \Vd0{$V_{_\mathrm{DS}}$}
\def \Vg0{$V_{_\mathrm{GS}}$}
\def \Vt0{$V_{_\mathrm{T}}$}
\def \Rc0{$R_{_\mathrm{C}}$}
\def \u0{$\mu_{_\mathrm{0}}$}

\newcommand{\til}{\raisebox{.1ex}{$\sim$}}	
\newcommand{\rpm}{\raisebox{.1ex}{$\pm$}}
\newcommand{\rp}{\raisebox{.1ex}{$+$}}
\newcommand{\gt}{\raisebox{.1ex}{$>$}}
\newcommand{\less}{\raisebox{.1ex}{$<$}}

\renewcommand{\thesection}{\arabic{section}}	% to number sections with Arabic numerals 1,2,3.. instead of default Roman numerals i,ii,iii.. 
\renewcommand{\figurename}{\textbf{\footnotesize{Figure}}}
\renewcommand{\thefigure}{\textbf{\footnotesize{\arabic{figure}}}}
\renewcommand{\bibfont}{\footnotesize}			% to modify the font size of bibliography entries 

\makeatletter
\renewcommand\p@figure{\arabic{figure}\expandafter\@gobble}	% To redefine the string used for cross-referencing figures without boldface.
\makeatother

\title{One-dimensional edge contacts to a monolayer semiconductor}

\author{Achint Jain}
\email{acjain@ethz.ch}
\affiliation{Photonics Laboratory, ETH Zürich, 8093 Zürich, Switzerland}

\author{Áron Szabó}
\affiliation {Integrated Systems Laboratory, ETH Zürich, 8092 Zürich, Switzerland}

\author{Markus Parzefall}
\affiliation{Photonics Laboratory, ETH Zürich, 8093 Zürich, Switzerland}

\author{Eric Bonvin}
\affiliation{Photonics Laboratory, ETH Zürich, 8093 Zürich, Switzerland}

\author{Takashi Taniguchi}
\author{Kenji Watanabe}
\affiliation {National Institute for Material Science, 1-1 Namiki, Tsukuba 305-0044, Japan}

\author{Palash Bharadwaj}
\affiliation{Department of Electrical and Computer Engineering, Rice University, Houston, TX 77005, USA}

\author{Mathieu Luisier}
\affiliation {Integrated Systems Laboratory, ETH Zürich, 8092 Zürich, Switzerland}

\author{Lukas Novotny}
\email{lnovotny@ethz.ch}
\affiliation{Photonics Laboratory, ETH Zürich, 8093 Zürich, Switzerland}

\date{\today}


\begin{abstract}
Integration of electrical contacts into van der Waals (vdW) heterostructures is critical for realizing electronic and optoelectronic functionalities. However, to date no scalable methodology for gaining electrical access to buried monolayer two-dimensional (2D) semiconductors exists. Here we report viable edge contact formation to hexagonal boron nitride (hBN) encapsulated monolayer \mos2. By combining reactive ion etching, \textit{in-situ} Ar$^+$ sputtering and annealing, we achieve a relatively low edge contact resistance, high mobility (up to \til\SI{30}{\centi\meter\squared\per\volt\per\second}) and high on-current density (\gt\SI[per-mode=symbol]{50}{\micro\ampere\per\micro\meter} at \Vd0\,=\,\SI{3}{\volt}), comparable to top contacts. Furthermore, the atomically smooth hBN environment also preserves the intrinsic \mos2 channel quality during fabrication, leading to a steep subthreshold swing of \SI{116}{\milli\volt}/dec with a negligible hysteresis. Hence, edge contacts are highly promising for large-scale practical implementation of encapsulated heterostructure devices, especially those involving air sensitive materials, and can be arbitrarily narrow, which opens the door to further shrinkage of 2D device footprint.
\end{abstract}

\maketitle

\global\small	% To reduce font size globally although it doesn't affect abstract, captions and titles. 

Two-dimensional electronic devices made from transition metal dichalcogenides (TMDCs) have gained prominence in recent years for next-generation integrated electronics \cite{Desai16} and nanophotonics applications. In particular, \mos2 combined with other 2D materials into vdW heterostructures, appears as an attractive candidate for future transistor architectures \cite{Iannaccone18}, atomically thin p-n junctions and tunnel diodes \cite{Frisenda2018b}, memristors \cite{Wang2018b}, high-efficiency photodetectors \cite{Bharadwaj15}, light emitting diodes \cite{Wang2018} and novel valleytronic devices \cite{Schaibley16}. Such heterostructures are often assembled in a top-down manner by picking-up discrete 2D material layers with a top hBN flake and placing the resulting stack on a target substrate. Although the presence of a top hBN layer on one hand serves to encapsulate the constituent 2D materials in the heterostructure, at the same time however, it also hinders the fabrication of direct electrical contacts to the underlying layers. Despite vdW heterostructures having been extensively investigated, a practical route for making electrical contacts to them in a scalable manner is still lacking. \\[-1.5ex] 

In TMDC heterostructures assembled without any encapsulation layer, further limitations arise when electrical contacts are made in a conventional top-contact geometry. In this scenario, the contact electrodes come in direct physical contact with a TMDC layer over a finite area. Since such a methodology inherently requires performing lithography on unprotected TMDC layers, it exposes them to foreign chemical species which are difficult to remove. Additionally, bare TMDC surfaces in air are susceptible to O$_2$ and H$_2$O adsorption \cite{Qiu12, Jariwala13, Park13}. Owing to their atomically thin nature however, mono- and few-layer TMDCs are quite sensitive to their immediate environment which includes both surface adsorbates from ambient exposure and processing residues. These act as unintentional dopants leading to a spatially inhomogeneous carrier density \cite{Wu16}, which causes device-to-device variations in threshold voltage \cite{Rahimi16, Smithe2017b} and Schottky barrier height \cite{Giannazzo15}. Besides doping, surface contaminants also scatter \cite{Ji2018} and trap charge carriers, thereby resulting in reduced mobility, low on-current \cite{Qiu12, Jariwala13, Park13}, increased flicker noise \cite{Sangwan13, Xie14}, hysteresis \cite{Shimazu16} and compromised optical properties \cite{Cho14}. Although measurements performed under high vacuum after \textit{in-situ} annealing have made it possible to observe the intrinsic electrical transport properties of \mos2 \cite{Baugher13, Smithe2017a}, unencapsulated devices measured in air show a drastic reduction in carrier mobility, implying that even short-term air exposure is detrimental for mono- and bi-layer \mos2 devices \cite{Qiu12, Jariwala13}.  \\[-1.5ex]

Therefore, for enabling a viable usage of TMDCs in integrated electronics, better contact techniques are needed that allow for encapsulation before contact patterning, in order to preserve the intrinsic material quality and achieve superior performance. Moreover, encapsulation is also essential for long-term ambient stability since it is well-known that most TMDCs, including \mos2, \mose2 and \ws2 undergo gradual oxidation in air at room temperature \cite{Peto2018}, which leads to further mobility degradation \cite{Lee15}, morphological changes \cite{Park16, Gioele16} and adversely affected photoluminescence \cite{Gao16}. In fact, some 2D materials like MoTe$_{2}$, HfSe$_{2}$, ZrSe$_{2}$, NbSe$_{2}$, black phosphorus and InSe, are so unstable in air that surface deterioration can be detected within a day \cite{Gioele16}. This restricts their assembly to an inert atmosphere \cite{Cao15} and encapsulation in hBN is commonly employed to limit air exposure \cite{Cao15, Lee15}. With such materials, lithographic contact fabrication prior to encapsulation is not only difficult but also impractical. \\[-1.5ex]

In order to circumvent the issue of making electrical contacts to hBN-TMDC-hBN heterostructures, a common practice is to embed additional layers of graphene \cite{Lee15, Cui15, Liu2015a} or metallic NbSe$_{2}$ \cite{Guan17, Sata17} within the stack to act as electrodes. Pre-patterning contact vias into the top hBN before pick-up \cite{Wang15b, Telford18} or transfer of pre-patterned metal films onto TMDCs \cite{Liu2018} (or vice-versa) have also been reported. However, alignment and transfer of multiple contact layers severely increases the fabrication complexity, especially in multilayer heterostructures, and becomes difficult to scale-up for practical purposes. Moreover, in case of graphene, the contact resistance (\Rc0) sensitively depends on the twist angle between the graphene and TMDC layers which poses further alignment challenges \cite{Liao2018}. Even though large-area chemical vapor deposition (CVD) growth of lateral graphene-\mos2 heterostructures has made progress in recent years \cite{Ling16, Guimaraes16, Zhao16}, hard to control growth inhomogeneities \cite{Zhao16, Suenaga2018} as well as ripples and strain induced by lattice mismatch still exist along 2D-2D edge interfaces \cite{Han2018}, which could ultimately hinder fabrication of very short channel (\less\SI{100}{\nano\meter}) devices. Another possibility is to fabricate tunneling contacts on encapsulated TMDCs \cite{Wang16, Cui17, Li17, Ghiasi2018}. However, such devices are restricted to very thin hBN (1-4L) or oxide (\til\SI{2}{\nano\meter}) \cite{Lee2016} encapsulation layers for optimum carrier injection. A more versatile approach is to etch through the top hBN layer in order to expose an edge of any buried 2D material of interest and form a one-dimensional (1D) `edge contact' to it \cite{Wang13, Karpiak2017}. Although such a strategy has been highly successful for graphene \cite{Wang13}, similar attempts to make 1D edge contacts to monolayer (1L) \mos2 \cite{Chai16, Moon17} and few-layer \wse2 \cite{Xu16} were met with limited success until now. \\[-1.5ex]

\begin{figure*}[t]
	\centering
	\includegraphics[width=1\textwidth]{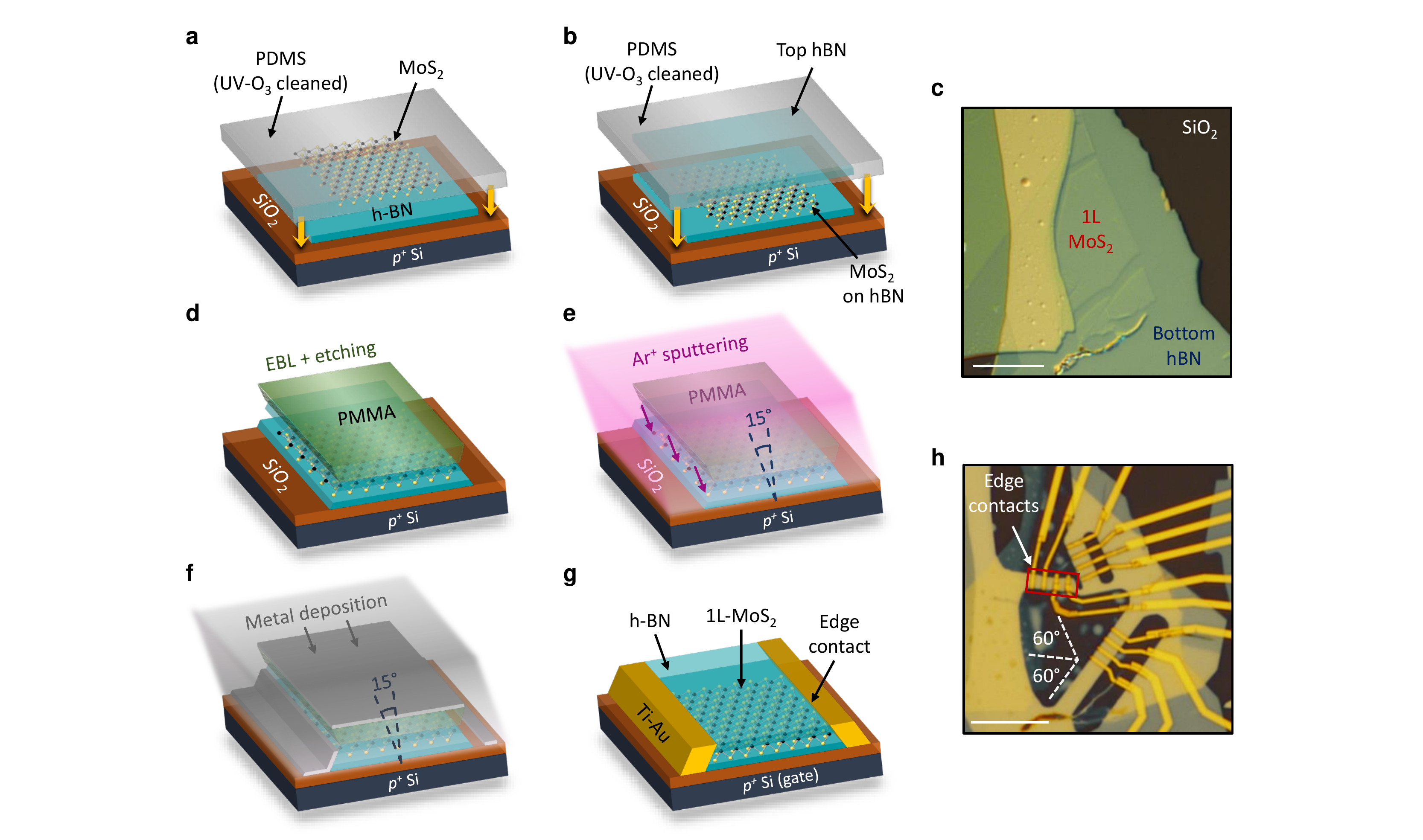}
	\caption{\label{fab}\footnotesize{\textbf{Fabrication of edge contacts. (a-b)} 3D illustration of the heterostructure assembly. 1L-\mos2 is exfoliated on PDMS and transferred onto an hBN layer. Subsequently, the \mos2 is fully encapsulated by stacking another hBN layer on top. \uvo3 cleaning of the PDMS surface before exfoliating \mos2 on it significantly reduces PDMS residues on \mos2 \cite{Jain2018}. \textbf{(c)} Differential interference contrast (DIC) optical image of a 1L-\mos2 flake transferred on hBN (\SI{24}{\nano\meter} thick) and vacuum annealed. \textbf{(d-g)} 3D illustration of edge contact fabrication. \textbf{(d)} The hBN-\mos2-hBN heterostructure is patterned by EBL and RIE to expose \mos2 edges. \textbf{(e)} Before metallization, \textit{in-situ} Ar$^+$ sputtering is done at \rp 15\textdegree\ and also -15\textdegree, inside a UHV chamber. This creates clean \mos2 contact edges by removing \moo3 and any adsorbed gas molecules. \textbf{(f-g)} Ti (5\,\rp\,\SI{5}{\nano\meter}) is then immediately deposited followed by Au (40\,\rp\,\SI{40}{\nano\meter}), both metals at \rp 15\textdegree\ as well as -15\textdegree, to form 1D edge contacts. \textbf{(h)}~Optical image of the \mos2 sample shown in \textbf{c}, contacted via edge contacts after hBN encapsulation. In the devices outlined in red, the \mos2 contact edges were not sputtered with Ar$^+$, to act as a control. Ar$^+$ treatment results in a lower contact resistance, as discussed later. The three sets of devices in \textbf{h} were aligned at 60\textdegree\ with respect to each other, in order to exclude any differences arising from the hexagonal crystal symmetry of hBN and allow for a more accurate comparison. For all devices, $L$ = \SI{1}{\micro\meter} and the contact length $L_\mathrm{c}$ = \SI{0.5}{\micro\meter}. Scale bars in \textbf{c}, \textbf{h}: \SI{10}{\micro\meter}.}}
\end{figure*}

Here we report reliable edge contact formation to hBN encapsulated 1L-\mos2. Our devices exhibit very low hysteresis together with a high mobility and steep subthreshold swing, highlighting the pristine interface quality achieved. By a systematic optimization of the fabrication process, we obtain a moderately low contact resistance and a high on-current (\gt\SI[per-mode=symbol]{50}{\micro\ampere \per\micro\meter}) with Ti-Au edge contacts, despite a vanishingly small contact area. The contact performance remains unchanged even at low temperatures, making edge contacts promising for cryogenic experiments and applications. Thus, our work introduces a universal approach for making efficient contacts to encapsulated 2D semiconductors, especially those sensitive to air, and marks an important step towards pristine devices with homogeneous electrical and optical characteristics on a macroscopic scale. We believe that with further improvement of the edge contact interface, by minimizing disorder and passivating in-gap edge states, as discussed later, even smaller \Rc0 is achievable. \\ [-1.5ex]

\section*{Edge contacts fabrication}

We will now discuss the fabrication strategy that we developed. Detailed process parameters can be found in Supporting Section S1. Bottom hBN flakes were exfoliated directly on \textit{p$^{+}$}Si/\sio2 (\SI{100}{\nano\meter}) substrates. 1L-\mos2 and top hBN flakes were separately exfoliated on GelPak\textsuperscript{\textregistered} PDMS (poly-dimethylsiloxane) stamps and transferred sequentially onto a suitable bottom hBN flake, as illustrated schematically in Figs.~\ref{fab}a,\,b. We found that PDMS can leave substantial residues behind after transfer which we minimized by pre-cleaning the PDMS surface in ultraviolet-ozone (\uvo3) prior to exfoliation (see Ref.~\onlinecite{Jain2018} for details). After each transfer, the resulting stack was annealed at \SI{200}{\degreeCelsius} in high-vacuum for \SI{3}{\hour} to release trapped bubbles, wrinkles and strain (if any) induced by PDMS during transfer \cite{Jain2018}. Figure~\ref{fab}c shows the optical image of a 1L-\mos2 flake transferred onto hBN from PDMS. To fully encapsulate the \mos2, another hBN flake was subsequently transferred on top. \\ [-1.5ex]

For device fabrication, bubble-free areas were chosen and patterned into rectangular sections by e-beam lithography (EBL) with PMMA (poly-methylmethacrylate) and reactive ion etching (RIE). Contact trenches were defined in a second EBL step and the exposed hBN-\mos2-hBN was etched away by RIE to create \mos2 edges for making contacts, as depicted in Fig.~\ref{fab}d (also see Supporting Fig.\ S1). The samples were then loaded into an e-beam evaporator for metal deposition, which we found to be the most critical part of the whole fabrication process. An etched \mos2 edge consists of dangling bonds as well as defects like Mo- and S-vacancies that are much more reactive than the basal plane of \mos2 \cite{Martincova2017}. During the time elapsed between etching and metal deposition, O$_2$ and H$_2$O molecules can not only bind to such edge sites but also potentially convert unpassivated Mo into \moo3 \cite{Martincova2017}. However, \moo3, which is often used as a hole transport layer in solar cells, hinders electron injection into \mos2 due to its high work-function \cite{Santosh2016}. This scenario is in strong contrast to top contacts where \moo3 formation is unlikely. \\ [-1.5ex]

Hence, immediately before metal deposition, \moo3 and any adsorbed O$_{2}$ or H$_{2}$O were removed by \textit{in-situ} Ar$^{+}$ sputtering at \rpm 15\textdegree\ tilt angle to expose a fresh \mos2 edge (Fig.~\ref{fab}e). Tilting the sample is necessary to access the etched hBN-\mos2-hBN sidewalls shadowed by an overhanging PMMA bilayer with an inward slope and avoid re-deposition of sputtered PMMA over the \mos2 edges. Ti-Au (5-\SI{40}{\nano\meter}) was then deposited at \rpm 15\textdegree\ tilt under a base pressure of \SI{1e-7}{\milli\bar} (Fig.~\ref{fab}f-g). After lift-off, the devices were annealed in Ar\,\rp\,H$_{2}$ at \SI{300}{\degreeCelsius} for 3 hrs to improve the Ti-\mos2 edge interface and reduce contact resistance (Supporting Section S2). Note that the use of Ti is essential for providing good adhesion to hBN sidewalls. Without Ti, pure Au tends to reflow and lose contact during annealing at \SI{300}{\degreeCelsius} (Supporting Section S4). The final set of devices with edge contacts are shown in Fig.~\ref{fab}h. \\ [-1.5ex]

\section*{Electrical characterization}

\begin{figure}[t]
	\centering
	\includegraphics[width=0.5\textwidth]{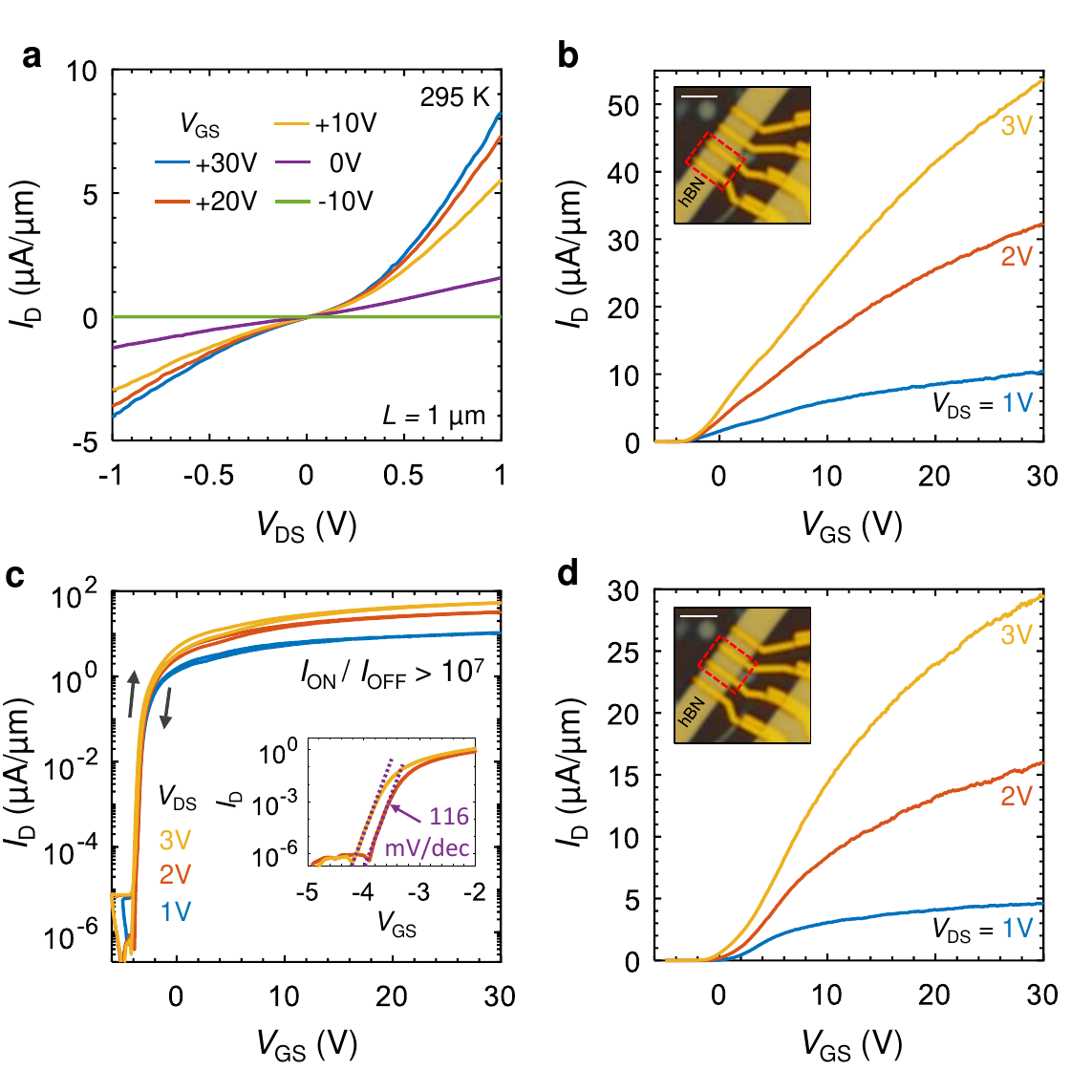}
	\caption{\label{IVchar}\footnotesize{\textbf{I-V measurements of monolayer \mos2 with edge contacts. (a)}~\Vg0 dependent two-probe \Id0-\Vd0 characteristics of a 1L-\mos2 transistor measured under ambient conditions. \textbf{(b)}~\Id0-\Vg0 characteristics of the same device demonstrating that edge contacts can support a high current density comparable to top contact devices with similar channel lengths but large metal-\mos2 overlap areas \cite{Liu2015b, Smithe2017a}. Inset: Optical image of the measured device (outlined). \textbf{(c)}~The data from \textbf{b} with \Id0 plotted on a log-scale showing both forward and backward sweeps to highlight the low hysteresis. Inset: Magnified plot of the subthreshold characteristics at \Vd0 = \SI{2}{\volt}, \SI{3}{\volt} (only forward sweeps) exhibiting a steep slope and \less \SI[per-mode=symbol]{1}{\pico\ampere\per\micro\meter} off current. \textbf{(d)}~\Id0-\Vg0 characteristics of the next device (inset) showing current magnitudes similar to the first device in \textbf{b}. The devices presented in these figures were etched with CHF$_{3}$\,\rp\,O$_{2}$ and sputtered with Ar$^+$ before metal deposition to form edge contacts. Scale bars in \textbf{b}, \textbf{d}: \SI{3}{\micro\meter}.}}
\end{figure}

Figure~\ref{IVchar}a shows the \Id0-\Vd0 output characteristics of an edge contacted 1L-\mos2 transistor exhibiting n-type behavior. A slight non-linearity at low \Vd0 indicates the presence of a small barrier at the contacts, as predicted by our quantum transport simulations (Supporting Section S6) and other computational studies \cite{Guo2015, Dong2019}. The \Id0-\Vg0 transfer characteristics of the same device are plotted in Figs.~\ref{IVchar}b-c on linear and log scales, respectively. A high current density reaching \SI[per-mode=symbol]{53.5}{\micro\ampere\per\micro\meter} at \Vd0 = \SI{3}{\volt} with an on-off ratio \gt $10^7$ can be observed. This clearly demonstrates that an efficient carrier injection is achievable via edge contacts, despite the lack of a 2D overlap between \mos2 and Ti. In Fig.~\ref{IVchar}c, each curve is comprised of both forward and backward sweeps which display a very small hysteresis. A magnified plot of the subthreshold characteristics is shown in the inset of Fig.~\ref{IVchar}c and reveals a low subthreshold swing (SS) of \SI{116}{\milli\volt}/dec maintained up to nearly 4 orders of magnitude. Realization of such a steep slope and low hysteresis was made possible here by encapsulation in hBN which not only protects the \mos2 channel from processing residues, but also provides an atomically smooth dielectric interface free of dangling bonds and defects. This significantly decreases the interface trap density in comparison with an exposed \mos2 layer on a \sio2 substrate. Moreover, the absence of thermally populated surface optical phonons in hBN at room temperature leads to a reduced scattering rate and enhanced carrier mobilities in \mos2 \cite{Dean10}. Using the relation \cite{Sze2006}, \\ [-3ex] 

\begin{equation}
\label{eqn0} \mathrm{SS} = (\ln\mathrm{10}) \frac{k_{_\mathrm{B}}T}{q} \left( 1 + \frac{C_{_\mathrm{it}}}{C_{_\mathrm{G}}}  \right) \\[2pt]
\end{equation}

\noindent
where $C_{_\mathrm{G}} = \SI{27.8}{\nano\farad\per\centi\meter\squared}$ is the gate capacitance per unit area and $C_{_\mathrm{it}} = q^{2}D_{_\mathrm{it}}$ is the interface capacitance per unit area, we estimated the density of interface trap states $D_{_\mathrm{it}}$ = \SI{1.7e11}{\per\electronvolt\per\centi\meter\squared}. This value is at least an order of magnitude lower than for unencapsulated, lithographically exposed \mos2 on \sio2 \cite{Zou2014, Choi2015} and ZrO$_2$ \cite{Desai16}. Note that in this sample, the SS is primarily limited by the back-gate dielectric thickness (\SI{100}{\nano\meter} \sio2 \rp\ \SI{24}{\nano\meter} hBN). From Eq.~\eqref{eqn0}, it is evident that for a larger gate capacitance (thinner dielectric), a smaller SS approaching the room temperature thermionic limit of \til\,\SI{60}{\milli\volt}/dec is anticipated. The \Id0-\Vg0 characteristics of a second device on the same sample are plotted in Fig.~\ref{IVchar}d, displaying a current density comparable to Fig.~\ref{IVchar}b. \\ [-1.5ex]

To investigate the influence of the metal-\mos2 contact edge cleanliness on carrier injection, a set of three control devices that were not  Ar$^+$ sputtered, were also fabricated on the same hBN-\mos2-hBN stack shown in Fig.~\ref{fab}h (outlined in red). All control devices conduct a significantly lower \Id0 than in Figs.~\ref{IVchar}b-d, indicating that carrier injection can be hindered if the \mos2 edge is not freshly cleaned immediately before metal deposition (Supporting Section S5). Additionally, it has been reported that Ti can partially oxidize during evaporation, depending on the vacuum level inside the deposition chamber, and thereby result in TiO$_\mathrm{x}$ formation at the contact interface \cite{McDonnell16}. To inhibit the oxidation of Ti, we deposited Ti-Au on another set of devices (again without Ar$^+$ sputtering) at a 10x lower base pressure of \til\SI{1e-8}{\milli\bar}, with negligible residual O$_2$ (Supporting Section S5). However, a low \Id0 is also observed in this case, revealing the existence of an \Rc0 dominated transport. This implies that a better vacuum does not lead to any appreciable change in the contact properties if Ar$^+$ sputtering is not done. It must be emphasized that an optimum post-deposition annealing temperature is also crucial for improving the contact interface (Supporting Section S2). Hence, our key finding here is that a clean \mos2 edge (before metallization) and annealing (after metallization) are both essential for forming good edge contacts, like those demonstrated in Figs.~\ref{IVchar}b-d. This likely explains why such a high current density had not been observed previously \cite{Chai16, Moon17}. \\ [-1.5ex]
 
\begin{figure*}[t]
	\centering
	\includegraphics[width=1\textwidth]{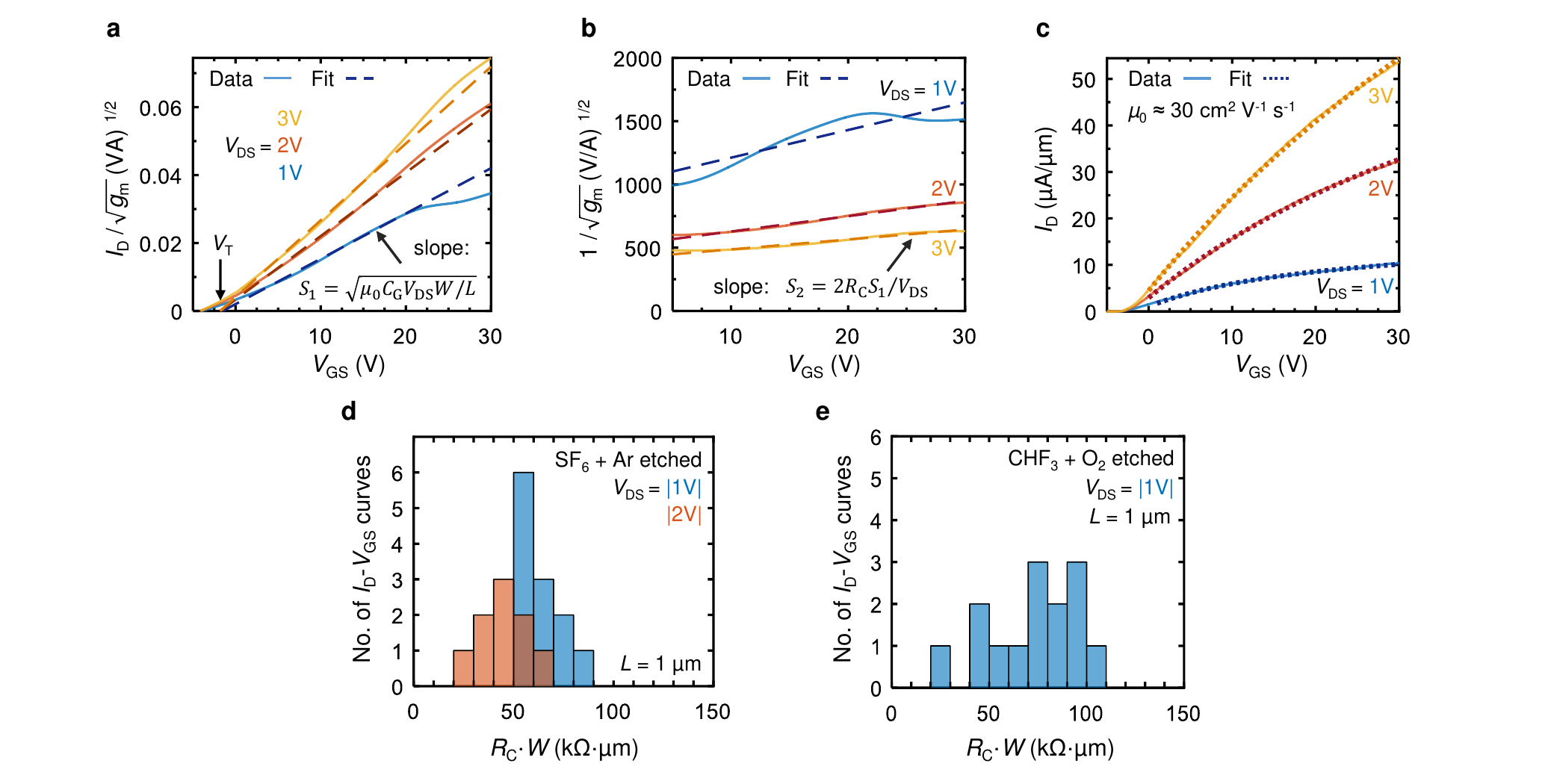}
	\caption{\label{rcmu}\footnotesize{\textbf{Mobility and contact resistance estimation. (a)} $Y$-function \textit{vs.}\ \Vg0 plot of the data shown in Fig.~\ref{IVchar}b. The dashed straight lines are fits to the linear region from which the mobility (\u0) and threshold voltage (\Vt0) can be extracted using the slope ($S_{{}_{\rm 1}}$) and x-intercept, respectively. \textbf{(b)} $1/\sqrt{g_\mathrm{m}}$ \textit{vs.}\ \Vg0 plot of the same dataset as \textbf{a}. The edge contact resistance can be estimated from the slopes ($S_{{}_{\rm 2}}$) of the linear fits using the given expression. \textbf{(c)} The \Id0-\Vg0 data from Fig.~\ref{IVchar}b fitted with the model in Eq.~\eqref{eqn3} choosing \u0, $\theta$ and \Vt0 as fit parameters. The goodness of the fit confirms that Eq.~\eqref{eqn3} can accurately model our \Id0-\Vg0 characteristics. \u0 $\approx$ \SI{30}{\centi\meter\squared\per\volt\per\second},  \Rc0$\cdot W$ = 27.8, 11.7 and \SI{8.3}{\kilo\Omega}$\cdot$\textmu m at \Vd0 = 1, 2 and \SI{3}{\volt} were estimated from the fits for this device. \textbf{(d)}~Histogram of \Rc0$\cdot$W values for six devices etched with SF$_{6}$\,\rp\,Ar. For each device, the \Rc0 was extracted by fitting the respective \Id0-\Vg0 curves with Eq.~\eqref{eqn3}. The histogram includes \Rc0 for both bias polarities (\Vd0 = \rp\SI{1}{\volt}, \SI{-1}{\volt}) owing to slightly asymmetric \Id0-\Vd0 curves (see Fig.~\ref{IVchar}a). In some cases, \Rc0 was also extracted from \Id0-\Vg0 curves recorded at \Vd0 = $\lvert\SI{2}{\volt}\rvert$ (orange bars). \textbf{(e)} Histogram of \Rc0$\cdot$W values at \Vd0 = $\lvert\SI{1}{\volt}\rvert$ for seven devices etched with CHF$_{3}$\,\rp\,O$_{2}$. An increased variability in \Rc0 can be seen with CHF$_{3}$\,\rp\,O$_{2}$ compared to SF$_{6}$\,\rp\,Ar.}}
\end{figure*}

Next, we want to characterize the intrinsic carrier mobility (\u0) and contact resistance (\Rc0) of our devices. For an ideal long-channel nMOSFET operating in the strong inversion regime, a linear dependence of \Id0 on \Vg0 is expected, given by \\ [-3ex]  

\begin{equation}
	\label{eqn1} I_{{}_\mathrm{D}} = \mu_{_\mathrm{0}}C_{{}_\mathrm{G}}\frac{W}{L}\,(V_{{}_\mathrm{GS, int}} - V_{{}_\mathrm{T}} - V_{{}_\mathrm{DS, int}}/\mathrm{2})\,V_{{}_\mathrm{DS, int}} \\[5pt]
\end{equation}

\noindent
where \Vt0 is the threshold voltage and the internal drain ($V_{{}_\mathrm{DS, int}}$) and gate ($V_{{}_\mathrm{GS, int}}$) voltages are equal to the externally applied bias. Typically, \u0 is extracted from the slope of linear \Id0-\Vg0 characteristics with the help of Eq.~\eqref{eqn1}. However, Fig.~\ref*{IVchar}b shows that \Id0 grows sub-linearly with \Vg0 for all \Vd0, which causes the mobility extracted in this manner to be underestimated. For a more accurate description of such \Id0-\Vg0 behavior, the presence of finite contact resistances \Rc0 in series with the \mos2 channel must be considered. In this scenario, the internal voltages seen by the channel get reduced to $V_{{}_\mathrm{DS, int}} = V_{{}_\mathrm{DS}} - \mathrm{2}I_{{}_\mathrm{D}}R_{{}_\mathrm{C}}$ and $V_{{}_\mathrm{GS, int}} = V_{{}_\mathrm{GS}} - I_{{}_\mathrm{D}}R_{{}_\mathrm{C}}$. Note that \Vt0 also gets modified to $V_{_{\mathrm{T}}} + {\delta}R_{{}_\mathrm{C}}V_{{}_\mathrm{GS}}$ but the drain-induced barrier lowering (DIBL) factor $\delta$ is small enough to be neglected in our long-channel devices. Equation~\eqref{eqn1} can then be re-written as \\ [-3ex] 

{\setlength{\abovedisplayskip}{-2pt}
\begin{equation}
	\label{eqn2} I_{{}_\mathrm{D}} = \mu_{_\mathrm{0}}C_{{}_\mathrm{G}}\frac{W}{L}\,\big[\,(V_{{}_\mathrm{GS}} - I_{{}_\mathrm{D}}R_{{}_\mathrm{C}}) - V_{{}_\mathrm{T}} - (V_{{}_\mathrm{DS}} - \mathrm{2}I_{{}_\mathrm{D}}R_{{}_\mathrm{C}})/\mathrm{2}\,\big]\,(V_{{}_\mathrm{DS}} - \mathrm{2}I_{{}_\mathrm{D}}R_{{}_\mathrm{C}}) \\[2pt]
\end{equation}
}
Rearranging Eq.~\eqref{eqn2} to solve for \Id0, we obtain \\ [-4ex]

\begin{gather}
	\label{eqn3} \implies I_{{}_\mathrm{D}} = \frac{\mu_{_\mathrm{0}}C_{{}_\mathrm{G}}}{\mathrm{1} + \theta(V_{{}_\mathrm{GS}} - V_{{}_\mathrm{T}} - V_{{}_\mathrm{DS}}/\mathrm{2})} \frac{W}{L}\,(V_{{}_\mathrm{GS}} - V_{{}_\mathrm{T}} - V_{{}_\mathrm{DS}}/\mathrm{2})\,V_{{}_\mathrm{DS}} \\[-1.5ex]
	\text{where} \quad \theta = \mathrm{2}R_{_\mathrm{C}}\,\mu_{_\mathrm{0}}C_{{}_\mathrm{G}}\frac{W}{L} \nonumber
\end{gather}

From Eq.~\eqref{eqn3}, we can infer that when $R_{_\mathrm{C}}\,\neq\,0$, \Id0 increases sub-linearly with \Vg0. To exclude the effect of \Rc0, one can first calculate ${\rm 1}/\sqrt{g_\mathrm{m}}$, as shown by Ghibaudo \cite{Ghibaudo1988} and Jain \cite{Jain1988}, where $g_\mathrm{m} \equiv \partial I_{{}_\mathrm{D}}/\partial V_{{}_\mathrm{GS}}$ is the transconductance of the device. \\ [-4ex]

\begin{equation}
\label{eqn4} \frac{\rm 1}{\sqrt{g_\mathrm{m}}} = \left(\frac{L}{\mu_{_\mathrm{0}}C_{{}_\mathrm{G}}V_{{}_\mathrm{DS}}W}\right)^{\rm 1/2} \big[\mathrm{1} + \theta(V_{{}_\mathrm{GS}} - V_{{}_\mathrm{T}} - V_{{}_\mathrm{DS}}/\mathrm{2})\big] \\[4pt]
\end{equation}

Upon multiplying Eqs.~\eqref{eqn3} and \eqref{eqn4}, $\theta$ can be eliminated and an expression commonly known as the $Y$-function is obtained, which depends linearly on \Vg0. \\ [-3.5ex]

\begin{equation}
	\label{eqn5} Y \equiv \frac{I_{{}_\mathrm{D}}}{\sqrt{g_\mathrm{m}}} = \left(\mu_{_\mathrm{0}}C_{{}_\mathrm{G}}V_{{}_\mathrm{DS}}\frac{W}{L}\right)^{\mathrm{1/2}} (V_{{}_\mathrm{GS}} - V_{{}_\mathrm{T}} - V_{{}_\mathrm{DS}}/\mathrm{2}) \\[4pt]
\end{equation}

By plotting $Y$ \textit{vs.}\ \Vg0 and using Eq.~\eqref{eqn5}, the mobility (\u0) and threshold voltage (\Vt0) can be extracted from the slope ($S_{_1}$) and x-intercept, respectively \cite{Smithe2017b, Chang2014}. Figure~\ref{rcmu}a is a plot of the $Y$-function for the data in Fig.~\ref{IVchar}b. It shows an approximately linear behaviour in the strong inversion regime from which a value of \u0 = \SI{29.2}{\centi\meter\squared\per\volt\per\second} can be derived. Lastly, we plot $1/\sqrt{g_\mathrm{m}}$ \textit{vs.}\ \Vg0 (Fig.~\ref{rcmu}b) and extract the slope ($S_{_2}$) of the linear region. \Rc0 can then be determined from the relation $R_{_\mathrm{C}} = S_{{}_{\rm 2}}V_{{}_\mathrm{DS}}/{\rm 2}S_{{}_{\rm 1}}$, derived using Eqs.~\eqref{eqn3}--\eqref{eqn5} \cite{Cho2018}. However, we found that due to random undulations of the derivative $\partial I_{{}_\mathrm{D}}/\partial V_{{}_\mathrm{GS}}$, the plots in Fig.~\ref{rcmu}b do not always remain linear in the entire $V_{{}_\mathrm{GS}} - V_{{}_\mathrm{T}} > V_{{}_\mathrm{DS}}/\rm{2}$ range for every device. As a consequence, the extracted slope can vary depending on the range chosen for the 1D polynomial fit. Hence, for a more reliable estimation of device parameters, we followed a slightly different approach and directly fitted our \Id0-\Vg0 curves with Eq.~\eqref{eqn3}, choosing all three unknowns (\Vt0, $\theta$ and \u0) as fitting parameters. We found this procedure to be more straightforward than the commonly used $Y$-function method. \\ [-1.5ex]

Figure~\ref{rcmu}c is a reproduction of the plots in Fig.~\ref{IVchar}b, fitted with Eq.~\eqref{eqn3} in the inversion regime for each \Vd0. The excellent quality of the fits indicates that the model in Eq.~\eqref{eqn3} describes our \Id0-\Vg0 characteristics very well. The estimated mobility \u0 = \SI{29.8}{\centi\meter\squared\per\volt\per\second} is also in good agreement with the value obtained from the $Y$-function plots in Fig.~\ref{rcmu}a. Knowing \u0 and $\theta$ from the fits, the contact resistance could then be deduced using Eq.~\eqref{eqn3} to be \Rc0$\cdot W$ = 27.8, 11.7 and \SI{8.3}{\kilo\Omega}$\cdot$\textmu m at \Vd0 = 1, 2 and \SI{3}{\volt}, respectively. It was found to decrease with increasing \Vd0 owing to enhanced Schottky barrier tunneling at higher bias voltages, as indicated by the non-linear \Id0-\Vd0 characteristics in Fig.~\ref{IVchar}a. Interestingly, these numbers are very similar to room temperature \Rc0 values reported for graphene top contacts on hBN encapsulated 1L-\mos2 devices (\SI{20}{\kilo\Omega}$\cdot$\textmu m) \cite{Cui15} as well as CVD grown lateral graphene\,-\,\mos2 contacts (10-\SI{50}{\kilo\Omega}$\cdot$\textmu m) \cite{Guimaraes16, Zhao16}. But at the same time, compared to the latter case, we observe a higher mobility owing to hBN encapsulation. These results unambiguously demonstrate that edge contacts can replace graphene contacts in encapsulated devices and achieve better performance with a less restrictive and scalable fabrication methodology. Moreover, use of graphene with \mos2 is currently limited to electron injection only, whereas with a proper choice of edge contact material, hole injection can also be feasible \cite{Guo2015}. \\ [-1.5ex]

\begin{figure*}[t]
	\centering
	\includegraphics[width=1\textwidth]{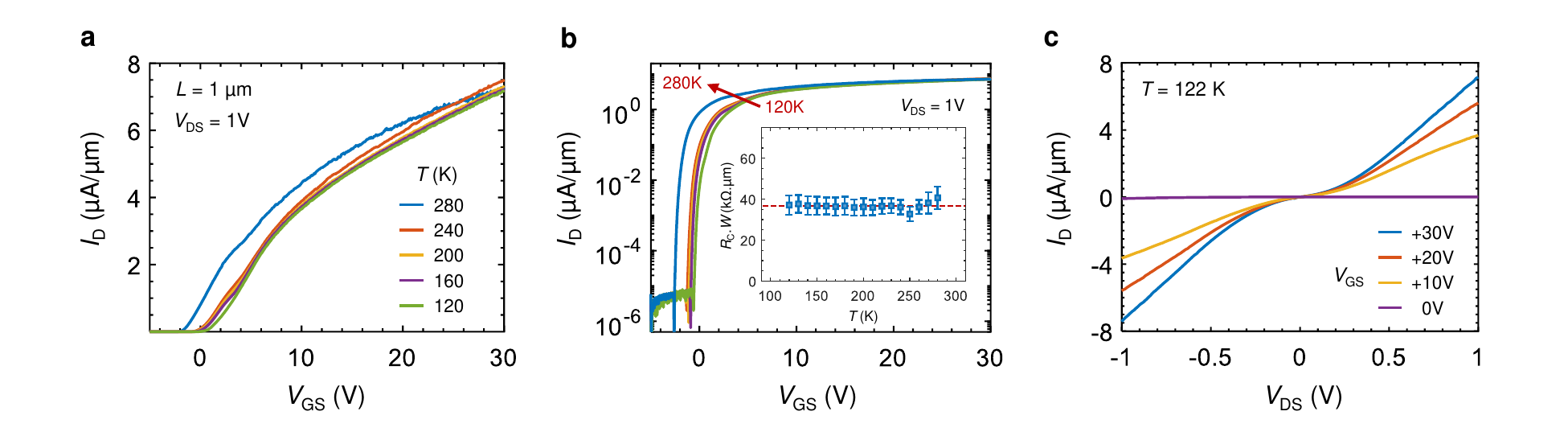}
	\caption{\label{lowT}\footnotesize{\textbf{Low temperature measurements.}~\Id0-\Vg0 characteristics of an edge contacted 1L-\mos2 transistor at various temperatures plotted on \textbf{(a)} linear and \textbf{(b)} log scales. Inset: Extracted \Rc0 values revealing a temperature independent behavior, which indicates that edge contacts can perform well even at low temperatures. The error bars denote 99\% confidence intervals of the fitted values. The horizontal red dashed line is a guide to the eye. \textbf{(c)}~\Id0-\Vd0 curves at \SI{122}{\kelvin} displaying similar characteristics as at room temperature (Fig.~\ref{IVchar}a)}. }
\end{figure*}

For completeness, it should be clarified that in our analysis \Rc0 is assumed to be independent of \Vg0 whereas in conventional contacts, it decreases asymptotically with increasing carrier density for low \Vg0 near the onset of inversion, and slowly saturates at high \Vg0. Such a behavior arises from the fact that in Ti-\mos2 top contacts with an interfacial oxide (often unintentional), increasing \Vg0 reduces the sheet resistivity of \mos2 below the contact region and also lowers the potential barrier for electron injection into \mos2 \cite{Szabo2019}, which increases the effective current transfer length $L_{_\mathrm{T}}$ \cite{Liu2014c}. Moreover, at the same time the applied \Vg0 also pushes the conduction band (CB) minimum closer to the metal Fermi level, thereby bending the CB more steeply near the contact edge, which narrows the effective Schottky barrier width \cite{Liu2014c}. This two-fold mechanism leads to a strong reduction of \Rc0 in top contacts as \Vg0 increases \cite{Allain2015}. However, in edge contacts where a 2D metal-\mos2 overlap region is absent ($L_{_\mathrm{T}} \approx 0$), the primary mechanism behind \Rc0 reduction with increasing \Vg0 is Schottky barrier narrowing, being more pronounced near the subthreshold region and saturating soon after. This causes \Rc0 to show a gate dependence that is weak enough to be neglected for $V_{_\mathrm{GS}} - V_{_\mathrm{T}} \gg V_{_\mathrm{DS}}/2 $, as substantiated by the constant slopes $S_{_1}$ and $S_{_2}$ in Figs.~\ref{rcmu}a-b for \Vg0 \gt\,\SI{5}{\volt}, which justifies our initial assumption. The model in Eq.~\eqref{eqn3} also fits well only in this regime. Hence, the \Rc0 we estimated is the \Vg0 independent value at large carrier densities, similar to Ref.~\onlinecite{Chang2014}. The most accurate way of extracting \Rc0 is the transfer length method (TLM). However, it requires fabrication of several devices with decreasing channel lengths and only works well when all devices have a very similar \Rc0 {\color{black}and \u0,} such that a plot of total resistance \textit{vs.}\ channel length follows a straight line. {\color{black}This has turned out to be challenging at present for our devices. Hence, we employed a simpler method, which gives a reasonable estimate of \Rc0 and \u0 for every single device and also helps in quantifying the device-to-device variability, unlike TLM.} \\ [-1.5ex]

Besides the device in Fig.~\ref{rcmu}c, we obtained similarly good fits for \Id0-\Vg0 curves measured from additional devices, which further corroborates the model we used (see Supporting Section S3 for more $I$-$V$ datasets). From these fits, an average \u0 = (20.5\,\rpm\,5.5)\,\si{\centi\meter\squared\per\volt\per\second} was found, where the error margin represents one standard deviation. To study the influence of etched hBN sidewall profiles on edge contacts, we tested two different hBN etch recipes \cite{Wang13, Autore2018}. Figures~\ref{rcmu}d-e are histograms of \Rc0 extracted from devices etched using the two recipes. For SF$_{6}$\,\rp\,Ar etched devices, we estimate an average \Rc0$\cdot W$ = (64.2\,\rpm\,9.6)\,\si{\kilo\Omega}$\cdot$\textmu m at \Vd0 = $\lvert\SI{1}{\volt}\rvert$ (blue bars). Since our \Id0-\Vd0 curves are slightly asymmetric in general (Fig.~\ref{IVchar}a), we extracted \Rc0 values from \Id0-\Vg0 fits for both positive and negative \Vd0. Some devices were also measured at \Vd0 = $\lvert\SI{2}{\volt}\rvert$ (orange bars) with an average \Rc0$\cdot W$ = (46\,\rpm\,10)\,\si{\kilo\Omega}$\cdot$\textmu m. In contrast to SF$_{6}$\,\rp\,Ar etched devices, those etched with CHF$_{3}$\,\rp\,O$_{2}$ show a wider distribution (Fig.~\ref{rcmu}e) and a higher mean value of \Rc0$\cdot W$ = (73.5\,\rpm\,23.4) \si{\kilo\Omega}$\cdot$\textmu m. We attribute this increased variability to greater etching inhomogeneity resulting from CHF$_{3}$\,\rp\,O$_{2}$ in comparison with SF$_{6}$\,\rp\,Ar, which we discovered upon scanning electron microscopy of bare hBN sidewalls (Supporting Fig.\ S1). \\ [-1.5ex] 

We further characterized another edge contacted 1L-\mos2 device at low temperatures inside a liquid nitrogen filled cryostat and is presented in Fig.~\ref{lowT}. Apart from the expected shift in threshold voltage \Vt0 to higher values \cite{Sze2006}, we find that the \Id0-\Vg0 characteristics as well as the edge contact resistance in Figs.~\ref{lowT}a-b remain essentially unchanged up to \SI{120}{\kelvin}. This observed temperature insensitivity of \Rc0 agrees very well with previous findings on 1D edge contacts to graphene \cite{Wang13} as well as CVD grown graphene edge contacts to 1L-\mos2 \cite{Guimaraes16}, and demonstrates that carrier injection into \mos2 via edge contacts occurs efficiently even under cryogenic conditions. Moreover, the \Id0-\Vd0 characteristics at \SI{122}{\kelvin} plotted in Fig.~\ref{lowT}c, behave similar to those at room temperature seen earlier for the device in Fig.~\ref{IVchar}a. To shed some light on this behavior, we performed \textit{ab initio} quantum transport simulations following the procedure described in our earlier publication \cite{Szabo2019} and are discussed in Supporting Section S6. In brief, the majority of the current injected via Ti edge contacts into \mos2 does not come from thermionic emission over the contact Schottky barrier, but rather tunneling across the barrier. Since the electron transmission probability near the Fermi level remains relatively constant as a function of energy (Supporting Fig.~S10), the tunneling current varies only weakly with temperature. We found that this tendency persists for a range of Schottky barrier heights that were evaluated. \\ [-1.5ex]  

\section*{Discussion and conclusions}

Strictly speaking, the true bandstructure of a semiconductor is defined for a lattice with an infinitely repeating unit cell. At \mos2 edges and grain boundaries, dangling bonds and Mo-, S-vacancies perturb the \mos2 bandstructure and give birth to additional localized `edge states' within the bandgap, as measured experimentally \cite{Bollinger2001, Wu16}. Such states were also observed in air-exposed \mos2 \cite{Wu16} and \wse2 \cite{Addou2018} devices, implying that adsorbed O$_2$ and H$_2$O do not fully passivate them. Passivation of dangling bonds and edge states is essential for good edge contacts \cite{Houssa2019}, which may be achieved by Ti-\mos2 bonding. However, if a van der Waals gap or trapped air molecules are present between the \mos2 edge and Ti, edge state passivation could be hindered, resulting in a high density of in-gap states at each electron injection site. By trapping incoming electrons, these states can cause a space charge region to build up which would repel further injected electrons. In this regard, \textit{in-situ} Ar$^+$ sputtering plays a key role in producing a clean \mos2 edge immediately before Ti deposition. Subsequent annealing at \SI{300}{\degreeCelsius} promotes atomic rearrangement and Ti-\mos2 bonding. The need for such extra measures does not arise in the case of edge contacts to graphene, where edge states (if any) are unable to trap carriers because of the absence of a bandgap. Even O$_2$ incorporation at the graphene edge was shown to have a negligible effect \cite{Wang13}, thus greatly simplifying fabrication of edge contacts to graphene. This scenario is fundamentally different from top contacts where the injected electrons do not encounter any edge states since the translational symmetry of the underlying \mos2 lattice is not broken (in the absence of interfacial reactions and defects) and on the contrary, a vdW gap is beneficial for avoiding Fermi level pinning \cite{Liu2018}. \\ [-1.5ex]

Interactions between the contact metal and \mos2 at the atomic scale and structural characteristics of the contact interface play a significant role in governing the performance of any contact. For edge contacts in particular, where carrier transfer is restricted to a single atomic edge, an optimum metal-\mos2 interface is crucial. This makes them more challenging to fabricate compared to top contacts which impose fewer constraints and can tolerate local non-idealities to a greater extent due to the availability of a finite area. Our main achievement here lies in the development of an optimized process for realizing low resistance edge contacts with a high density of current injection per atomic site. Further studies are needed nevertheless to unravel the rich physics and chemistry occurring at the contact interface. {\color{black}Atomically resolved cross-sectional transmission electron microscope (TEM) imaging can be performed to gain better insights into the contact morphology, interface quality and atomic configuration of edge contacts. This would lead to a deeper understanding of the transport behavior and provide valuable guidelines for further improvement of the contact performance.} It is possible that unpassivated edge states at interface voids still undermine the performance of our devices \cite{Houssa2019} and also cause undesired Fermi level pinning \cite{Chen2017}. Suitable chemical termination of dangling bonds could be a promising strategy to passivate edge states, de-pin the metal Fermi level and reduce \Rc0 even further. Apart from \mos2, air sensitive TMDCs like HfS$_2$, ZrS$_2$, etc.\ where edge states are expected to lie at shallow levels close to the band extrema, which makes them more immune to defects, appear as attractive materials for edge contacts \cite{Pandey2016}. \\ [-1.5ex]

It should be emphasized that even though \til\si{\micro\meter} long Ti-Au \cite{Liu2015b}, Ag-Au \cite{Smithe2017b} and In-Au \cite{Wang2019} top contacts on 1L-\mos2 have resulted in a lower \Rc0 than that obtained in this work, in order to be fair, a comparison should be made with top contacts scaled down to sub-nm overlap lengths. However, it has been shown that the \Rc0 begins to increase considerably for contact lengths smaller than the current transfer length in both mono-\cite{Liu2014c} and multi-layer \mos2 \cite{English2016}. This implies that conventional contacts cannot be scaled down beyond a certain limit, thereby restricting the minimum achievable device footprint (gate length + 2 x contact length). To ensure scaling of TMDC based devices, scalable contact geometries that work efficiently irrespective of dimensions are necessary. This bottleneck could be overcome by means of edge contacts, which do not require a 2D overlap with TMDCs and thus, in principle, can be made as narrow as possible. Another domain where edge contacts can outperform top contacts is multilayer TMDCs, in which carrier injection only via the topmost layer suffers from added interlayer hopping resistances that limit the current transport to top few layers \cite{Das2013}, whereas with edge contacts, each layer can be individually contacted for achieving higher current densities \cite{Schulman2018}. In this regard, 1T-phase edge contacts to few layer \mos2 \cite{Kappera2014} and \mote2 \cite{Sung2017} also seem to be an attractive choice, although inducing a 2H $\rightarrow$ 1T phase transition under the contact regions after encapsulation can be problematic. \\ [-1.5ex]
 
Lastly, the possibility to encapsulate 2D materials before processing with chemicals remains the biggest advantage of edge contacts for building clean devices. Fundamental studies rely on high interface quality and macroscopic homogeneity for uncovering new physical phenomena, which can benefit from edge contacts fabricated after encapsulation. Edge contacts are especially promising for 2D materials unstable in air for which fabrication of top contacts is challenging due to restrictions imposed by encapsulation inside an inert atmosphere before being exposed to air. Often such heterostructures are built in a top-down manner and the need to make contacts to buried layers demands pick-up of additional graphene sheets. In such scenarios, edge contacts provide a much higher flexibility in heterostructure assembly and can be scaled-up to integrated circuits employing multiple metal layers separated by insulating dielectric layers. Thus, we envision that edge contacts will bring devices based on 2D materials one step closer to practical implementation and open up new pathways in 2D materials research. \\ [-1.5ex]

\phantomsection
\section*{Associated Content}
\textbf{Supporting Information.} Detailed description of the fabrication procedure, plots showing the influence of annealing, data from additional devices, pure Au edge contacts without Ti, edge contacts without Ar$^+$ sputtering, quantum transport simulations. \\ [-1.5ex]

\section*{Methods} \label{methods}
\textbf{Edge contact fabrication.} See Supporting Section S1 for a step-by-step process flow. \\ [-1.5ex]

\textbf{Electrical characterization.} I-V measurements were carried out using a Keithley 2602B source meter in two-probe configuration. All devices were measured in air at room temperature (except those shown in Fig.~\ref{lowT}). For calculating $g_{m}$, the \Id0-\Vg0 curves were smoothened by cubic spline interpolation in MATLAB to reduce the noise before differentiation. \\ [-1.5ex]

\footnotesize
\section*{Acknowledgments}
This research was supported by the Swiss National Science Foundation (grant no.\ 200021\_165841), ETH Zürich (ETH-32 15-1) and CSCS (Project s876). Use of the cleanroom facilities at the FIRST Center for Micro and Nanoscience, ETH Zürich is gratefully acknowledged. TT and KW acknowledge support from the Elemental Strategy Initiative conducted by the MEXT, Japan and JSPS KAKENHI (grant no.\ JP15K21722). AJ would like to thank Aroosa Ijaz for invaluable help during sample fabrication. \\[-1.5ex] 

\section*{Author Contributions}
ML, AJ and LN conceived the project. AJ developed the fabrication procedure, carried out the measurements and analyzed the experimental data. \'AS and ML performed the quantum transport simulations. MP built the electrical characterization setup, wrote the LabVIEW scripts for recording I-V data and provided experimental support at various stages. Low temperature transport measurements were performed together with EB. TT and KW synthesized the hBN crystals used in this study. LN, ML and PB supervised the project. AJ wrote the manuscript with inputs from MP, ML and LN. \\[-1.5ex] 

% Bibliography

\onecolumngrid 	% For balancing the bibliography columns with RevTex



\section*{Supplementary material}

# *Supporting Information*

# One-dimensional edge contacts to a monolayer semiconductor

Achint Jain,[1, *] Áron Szabó,[2] Markus Parzefall,[1] Eric Bonvin,[1] Takashi Taniguchi,[3]
Kenji Watanabe,[3] Palash Bharadwaj,[4] Mathieu Luisier,[2] and Lukas Novotny[1, †]

[1]*Photonics Laboratory, ETH Zürich, 8093 Zürich, Switzerland*
[2]*Integrated Systems Laboratory, ETH Zürich, 8092 Zürich, Switzerland*
[3]*National Institute for Material Science, 1-1 Namiki, Tsukuba 305-0044, Japan*
[4]*Department of Electrical and Computer Engineering, Rice University, Houston, TX 77005, USA*

# List of Figures

# S1. Fabrication details

**hBN-MoS$_2$-hBN heterostructure assembly**: Bottom hBN flakes were directly exfoliated on O$_2$ plasma cleaned $p^+$Si/SiO$_2$ (100 nm) substrates using a blue tape (Nitto). Naturally occurring MoS$_2$ crystals (SPI Supplies) were exfoliated on viscoelastic PDMS stamps (Gel-Film® PF-40-X4 sold by Gel-Pak®). We found that PDMS often has a significant amount of uncrosslinked (or 'loosely bound') dimethylsiloxane oligomers on its surface[1] that can contaminate 2D materials exfoliated on PDMS[2]. In order to achieve high carrier mobilities and avoid unintentional doping, it is crucial to minimize PDMS residues and transfer MoS$_2$ in a pristine manner. Therefore, prior to MoS$_2$ exfoliation, all PDMS stamps were treated with UV-O$_3$ in a Bioforce Nanosciences UV-ozone ProCleaner for 30 min (manufacturer specified illumination intensity: 14.76 mW/cm$^2$), following the procedure outlined in ref. [2]. UV-O$_3$ exposure breaks down unwanted surface oligomers and forms a residue free, few ∼ nm thin SiO$_x$ layer on the PDMS surface[3]. After UV-O$_3$ treatment, PDMS stamps were left in ambient air for 2 h to deactivate the surface termination and cause a partial hydrophobic recovery of the PDMS surface. This wait interval prevents bonding between PDMS and the blue tape and helps to increase the yield during exfoliation.

Bulk MoS$_2$ crystals were then exfoliated on the clean PDMS stamps after 2 h and monolayer flakes were identified in an optical microscope. For transfer, PDMS stamps with MoS$_2$ were placed on a transparent quartz plate and aligned on top of suitable bottom hBN flakes on SiO$_2$ using a SÜSS MicroTec MJB4 mask aligner. All transfers were carried out in air. Upon coming in contact, the hBN-MoS$_2$ stacks were heated to ∼65 °C for 2 min with a Peltier module kept underneath the Si/SiO$_2$ substrates. After allowing for a few minutes to cool down, the PDMS stamps were then slowly detached. The transferred MoS$_2$ flakes on hBN were annealed at 200 °C for 3 h in high vacuum ($<1 \times 10^{-5}$ mbar) to remove any remaining PDMS residues and release transfer induced compressive strain as well as accumulated bubbles/wrinkles in MoS$_2$[2]. The same procedure was followed to transfer the top hBN flakes onto the MoS$_2$ - (bottom) hBN stacks and the resulting heterostructures were again vacuum annealed at 200 °C for 3 h to reduce the density of bubbles.

**Electron beam lithography (EBL) and etching**: Clean, bubble-free areas in the hBN-MoS$_2$-hBN heterostructures were chosen and patterned into well-defined rectangular channels by EBL (RAITH150 Two) with a bilayer of PMMA 50k (4% in chlorobenzene) and 950k (4.5% in anisole). Each layer was spin-coated at 5000 rpm for 45 s and baked at 180 °C for 4 min. After EBL (electron dose: 400 μC/cm$^2$ at 30 kV), PMMA was developed in MIBK:IPA (1:3) for 60 s and the exposed areas were etched away by reactive ion etching (RIE) in an Oxford Plasmalab 80 Plus system. For making edge contacts into the rectangular channels, EBL and RIE were repeated again. RIE was performed with either **(a)** CHF$_3$ + O$_2$ plasma (40 + 4 sccm, 50 W power and 37.5 mTorr pressure, etch rate: ∼36 nm/min)[4] or **(b)** SF$_6$ + Ar plasma (20 + 20 sccm, 50 W power and 100 mTorr pressure, etch rate: >90 nm/min)[5]. Scanning electron microscope (SEM) images of the hBN surfaces resulting from the two etching recipes are shown in Fig. S1 for comparison and reveal some striking differences. Reactive ion etching of hBN with CHF$_3$ + O$_2$ proceeds primarily via chemical interactions that strongly favor etching along certain in-plane crystal directions. It can be noticed in Fig. S1a that this highly anisotropic etch rate leaves behind a high density of hBN pyramids in the etched regions and forms sidewalls with very narrow triangular crevices (red dotted lines). It is reasonable to assume that during metal deposition over such hBN sidewalls, Ti grains might not fill these fine crevices entirely, resulting in a loss of contact with the MoS$_2$ edge at certain spots.

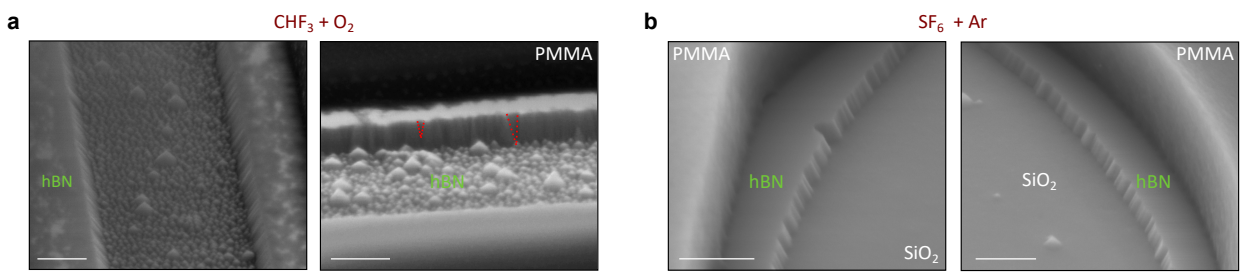

**Figure S1**. **Etched hBN roughness.** **(a)** Tilted view SEM images of contact trenches in a test hBN flake, etched half-way through with CHF$_3$ + O$_2$ for 210 s. The etched regions display a very rough bottom surface, with partially etched hBN pyramids, and sidewalls featuring narrow, vertical crevices with triangular facets, arising from the crystallographic planes of hBN. Two such crevices have been indicated by red dotted lines. **(b)** SEM images of a curved trench etched in another (thinner) hBN flake with SF$_6$ + Ar for 20 s. It exhibits a smooth bottom SiO$_2$ surface with a significantly reduced density of unetched hBN pyramids. No triangular crevices like those indicated in **a** could be resolved in this flake, although considerable sidewall roughness is still present. All images were taken immediately after etching, without removing PMMA, in order to avoid deposition of any organic residues that might smear out the sharp hBN features and reduce image contrast. Upon interaction with the electron beam (5 keV), the PMMA layer retracted away from the sides, allowing hBN to be imaged. Scale bars: 200 nm in all images.

Moreover, within a single device, source and drain contacts formed along parallel etched trenches would have different faceting due to the 60° rotational symmetry of hBN (and not 90°), giving rise to asymmetric $I_D$-$V_{DS}$ characteristics. In case of $SF_6$ + Ar (1:1), the introduction of Ar adds a physical sputtering contribution to the etch process. This leads to more isotropic in-plane etching, resulting in sidewalls devoid of triangular facets, as shown in Fig. S1b and lower contact variability (see Fig. S6).

**Metal deposition and annealing**: This is the most critical part of our fabrication process. As discussed in the main text, etched $MoS_2$ edges have a large number of dangling bonds that can host in-gap edge states[6] and act as adsorption sites for air molecules ($O_2$, $H_2O$). These adsorbed species could likely hinder covalent bonding between the contact metal (Ti) and $MoS_2$, leaving edge states unpassivated, which subsequently act as traps for injected carriers. Moreover, Mo atoms located at the edge can oxidize to form $MoO_x$ which, owing to its high work-function, would pose a further barrier for electron injection[7]. Therefore, in order to limit $MoS_2$ oxidation after RIE, the samples were immediately loaded into an electron beam evaporator (Plassys MEB550S) for metal deposition, with only a few minutes of air exposure in between. While loading, the etched contact trenches were aligned parallel to the axis of the tilting motor and thereafter, the samples were not rotated in-plane at any point during sputtering and evaporation. To remove any $MoO_x$ and form a clean $MoS_2$ edge just before metal deposition, the contact edges were sputtered *in-situ* with an $Ar^+$ ion beam at +15° and -15° tilt (Fig. S2b) for 15 s each (3.5 sccm flow rate, 250 V beam voltage, 50 V acceleration voltage, 10 mA beam current). Since ion-gun parameters can vary from machine-to-machine, in order to make it easier for others to reproduce our recipe, we estimated the etch rate of bare $MoS_2$ in our case by comparing the optical images in Fig. S3 before and after sputtering.

Ti (5 nm) was then deposited at 0.2 nm/s rate under a base pressure of $1 \times 10^{-7}$ mbar, first at +15° tilt and then 5 nm again at -15° tilt, as depicted in Fig. S2c. Tilting was necessary to avoid shadowing of the $MoS_2$ edge by the overhanging bilayer PMMA sidewalls. In this manner, a nearly conformal Ti layer could be deposited over the hBN sidewalls despite the rough topography, as visible in Fig. S2d. Au (40 + 40 nm) was then deposited at 0.2 nm/s in the same way. Note that the choice of deposition angle was constrained by the PMMA undercut angle in our case. Deposition at 20° caused the PMMA sidewalls to be also partially coated with Ti-Au, leaving behind vertical ears upon lift-off, while deposition at 30° resulted in no lift-off. After lift-off in hot acetone, the samples were annealed at 300 °C in Ar + $H_2$ (380 + 20 sccm) for 3 h inside a quartz tube furnace to improve the contacts. Finally, for electrical characterization, wire-bonds were made manually with a 25 μm diameter tungsten wire and silver epoxy (CircuitWorks® CW2400). The epoxy was cured at 80 °C for 30 min in a vacuum oven. Unlike conventional ultrasonic wire-bonding, Ag epoxy helped to prevent shorting with the Si back-gate through the 100 nm $SiO_2$ layer.

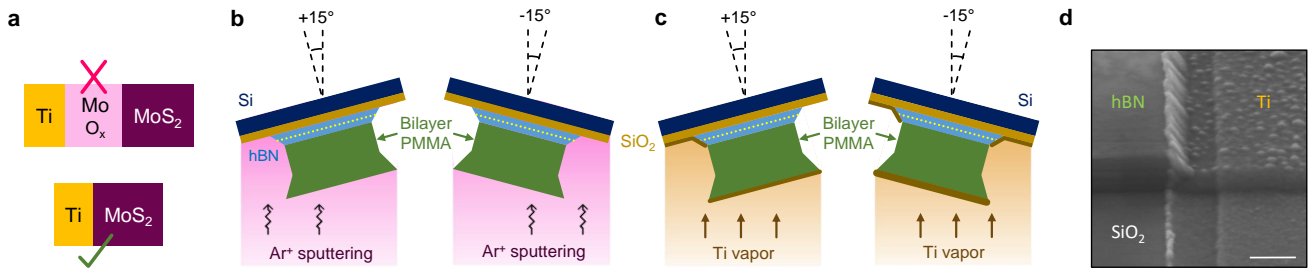

**Figure S2**. **Edge contact metallization.** **(a)** Sketch showing possible existence of $MoO_x$ at the Ti-$MoS_2$ interface (top) whereas a $MoO_x$-free interface is desired (bottom). **(b)** Schematic illustration of an hBN-$MoS_2$-hBN heterostructure mounted upside-down inside an e-beam evaporator chamber, undergoing *in-situ* $Ar^+$ sputtering at +15° and -15° tilt to remove $MoO_x$ (if any) and absorbed $O_2$, $H_2O$ molecules from both $MoS_2$ contact edges before metal deposition. **(c)** Schematic illustration of Ti deposition at +15° and -15° tilt, immediately after $Ar^+$ sputtering. **(d)** SEM image of a test hBN flake featuring a conformal deposition of Ti over the hBN sidewall, despite its roughness. Scale bar: 100 nm.

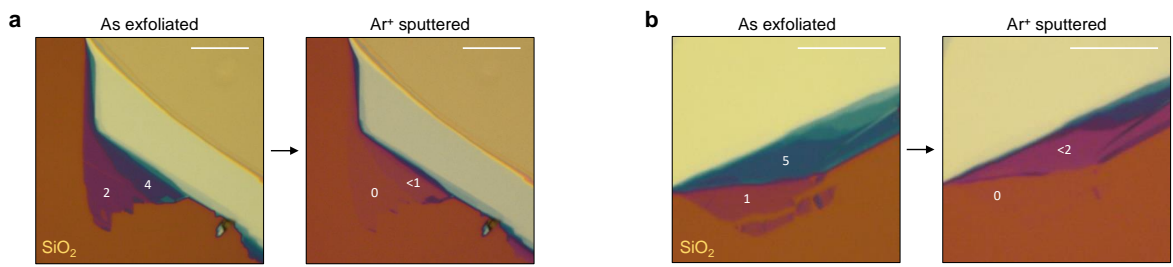

**Figure S3**. **$MoS_2$ etch rate characterization.** Optical images of $MoS_2$ flakes, as exfoliated on Si/$SiO_2$ (285 nm) substrates, and after $Ar^+$ sputtering for 15 s (without tilting). The number of layers, before and after sputtering, have been indicated in all images from which an etch depth of ~3.5 layers in 15 s can be estimated. Scale bars: 10 μm in all images.

## S2. Significance of annealing

Post metal-deposition annealing is commonly employed to lower the contact resistance ($R_C$) in 2D material devices. We tested several temperatures for this purpose and found that for edge contacts annealed in Ar + H$_2$ (380 + 20 sccm), $I_D$ improved remarkably with increasing temperatures up to 400 °C, as shown in Fig. S4a. At the same time, the high interface quality of our devices remained preserved due to hBN encapsulation, as evidenced by the unchanged steep subthreshold slope and negligible hysteresis in all $I_D$-$V_{GS}$ plots in Fig. S4b. Besides this, we also varied the annealing time at a fixed temperature but did not notice a substantial change in $I_D$ after a total duration >1.5 h (Fig. S4c). We further extracted the $R_C$ and mobility values from the curves in Fig. S4a and are plotted in Figs. S4d-e, respectively. A reduction in $R_C$ and mobility enhancement at higher temperatures clearly highlight the importance of annealing. However, we also observed that too high temperatures can affect the long-term ambient stability of our devices to a certain extent. Considering these facts, a temperature of 300 °C and time 2-3 h were therefore chosen as optimum for our fabrication recipe. Lastly, un-annealed devices measured directly after metal deposition and lift-off, can sometimes exhibit a large hysteresis in the transfer characteristics, as seen in Fig. S5a. Annealing can help to get rid of such hysteresis (Fig. S5b).

Thus, we can conclude that annealing is indispensable for edge contacts. In addition to Ar + H$_2$, we also tested annealing of edge contacts in a high vacuum environment ($<1 \times 10^{-5}$ mbar), but the former resulted in a slightly better performance. Several processes can occur simultaneously during annealing: metal-MoS$_2$ covalent bonding, reduction of interfacial oxides and sulfur atoms by H$_2$ molecules, inter-diffusion of edge atoms (metal into MoS$_2$ or vice-versa), defect migration and edge reconstruction. Presently, it is unclear what processes dominate in case of edge contacts and are responsible for the reduction of $R_C$ upon annealing, although Fig. S4c points towards a weaker contribution of diffusion, which is a slow, time-dependent process. Understanding the various mechanisms at play and developing more effective annealing procedures is a promising subject for future studies.

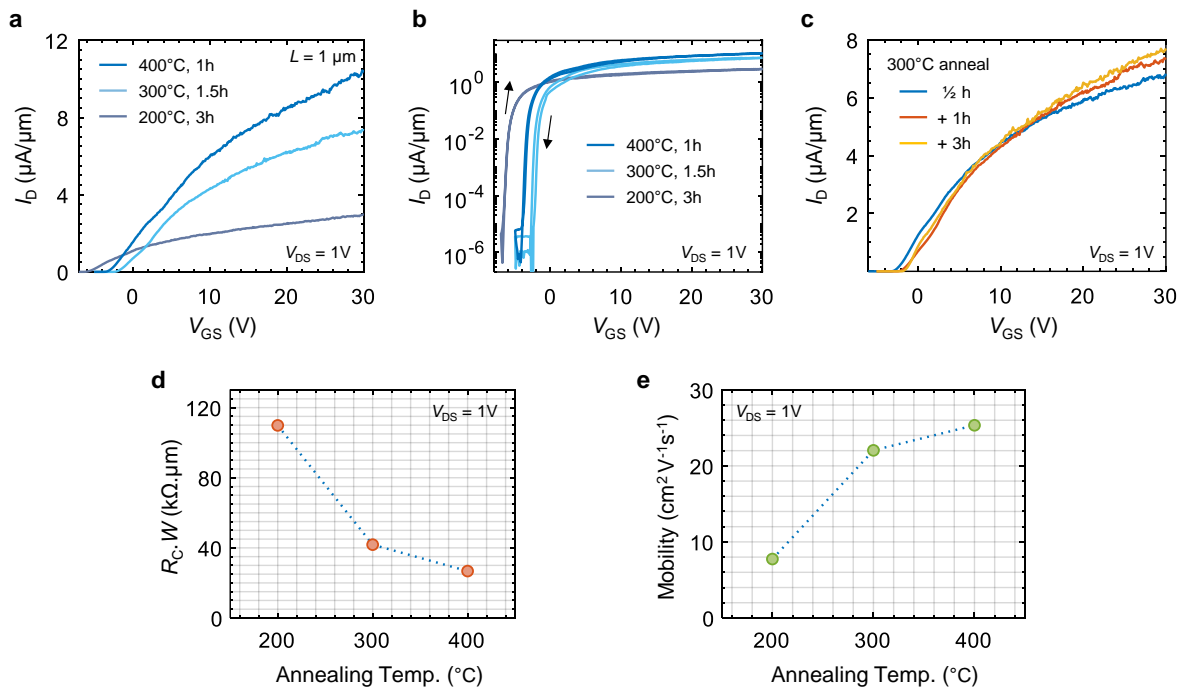

**Figure S4**. **Contact resistance reduction by annealing.** (a) $I_D$-$V_{GS}$ characteristics of a 1L-MoS$_2$ device measured after annealing in Ar + H$_2$ at three different temperatures. It can be seen that the current density increases monotonically with the annealing temperature, implying a reduction in $R_C$ upon annealing. This enhancement can possibly be attributed to improvement of the edge contact interface, leading to reduced scattering/trapping of electrons and thus, more efficient current injection. (b) $I_D$-$V_{GS}$ data shown in **a**, plotted on a log-scale, exhibiting a consistently steep subthreshold slope after each annealing cycle. Moreover, all plots comprise of both forward and backward sweeps that display very small hysteresis (sweep directions marked by arrows). This indicates that in our devices, the MoS$_2$ crystal quality and the low interface trap density (~$10^{11}$ eV$^{-1}$ cm$^{-2}$), remain preserved by hBN encapsulation and do not deteriorate, at least up to 400 °C. (c) $I_D$-$V_{GS}$ characteristics of the same device after annealing at 300 °C for 30 min, 1 h and 3 h, incrementally. These results reveal that prolonged annealing (>1.5 h), at a fixed temperature, does not cause any significant change in the contact properties. (d-e) Contact resistance and mobility values extracted from the plots in **a** for each annealing temperature. Note: All data presented here was recorded from the same device shown in Figs. 2a-c of the main text. After each annealing cycle, the sample was cooled down and measured at room temperature in air.

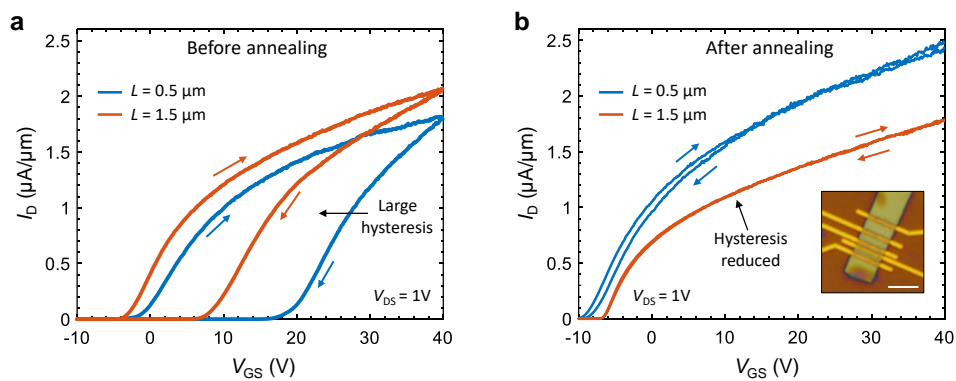

**Figure S5. Hysteresis reduction by annealing.** $I_D$-$V_{GS}$ transfer characteristics of 1L-MoS$_2$ devices with Ti-Au (10-60 nm) edge contacts, **(a)** as-fabricated and **(b)** after annealing in Ar + H$_2$ at 200 °C for 3 h. The as-fabricated devices exhibit a large hysteresis between the forward and reverse sweeps, which nearly vanishes upon annealing, accompanied by a shift in the threshold voltages to lower values. The gate voltage sweep directions have been indicated by arrows. Inset: Optical image of the measured devices. In this sample, the SiO$_2$ thickness was 285 nm and no *in-situ* Ar$^+$ sputtering was performed (which explains the lower $I_D$ compared to Fig. S4). Scale bar: 4 µm.

## S3. Data from additional devices

Here we show I-V data from devices in which edge contacts were etched with $SF_6$ + Ar. The $MoS_2$ flake shown in Fig. S6a was exfoliated on PDMS and transferred to hBN (Fig. S6b-c). After encapsulating the $MoS_2$ with another hBN and patterning the resulting stack into two rectangular segments, edge contacts were fabricated on each (Fig. S6d). The $I_D$-$V_{DS}$ characteristics of one such device are plotted in Fig. S6e and display a nearly linear behavior without any sign of saturation, at least until 3 V. In Fig. S6f, the $I_D$-$V_{GS}$ curves of two representative devices are shown and exhibit very similar characteristics, indicating a low-variability in contacts etched with $SF_6$ + Ar, as discussed in Section S1 above. All $I_D$-$V_{GS}$ curves in Fig. S6f have been fitted with the model used in the main text. The excellent quality of the fits further corroborates the validity of the model. Log-scale $I_D$-$V_{GS}$ plots are shown in Fig. S6g, revealing a steep subthreshold slope and high on/off ratio, similar to Fig. 2c of the main text. These measurements, together with those in the main text, demonstrate that with our fabrication procedure, edge contacts can be made reproducibly.

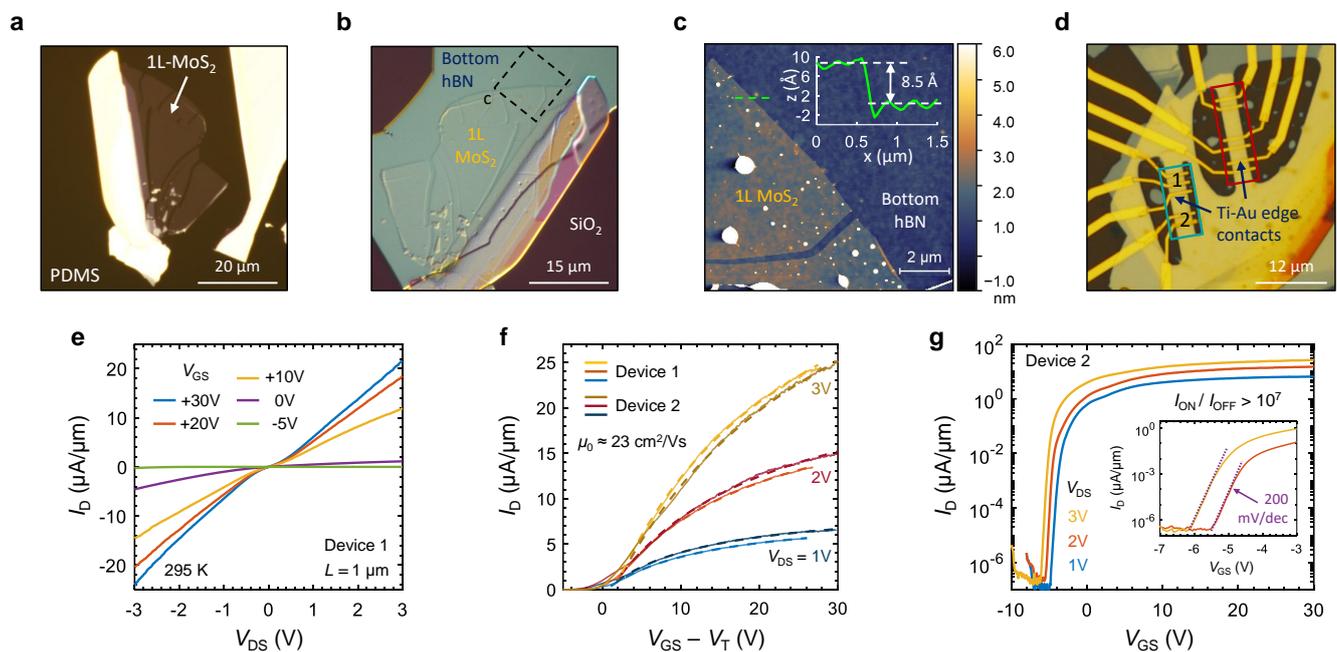

**Figure S6**. **$SF_6$ + Ar etched devices.** (a) Optical image of a 1L-$MoS_2$ flake exfoliated on UV-$O_3$-cleaned PDMS. (b) Differential interference contrast (DIC) image of the same flake after having been transferred to hBN (25 nm thick) and vacuum annealed. (c) AFM topography map of the 10 µm × 10 µm region outlined in **b**, displaying a pristine $MoS_2$ surface with few interfacial bubbles (bright spots). Inset: Cross-section profile along the green dashed line, revealing a thickness close to monolayer. (d) Optical image of six devices built using the $MoS_2$ flake in **b** after hBN encapsulation. In all devices, $L$ = 1 µm, $W$ = 3 µm and the contact length $L_C$ = 0.5 µm. The edge contacts outlined in blue were etched by RIE with $SF_6$ + Ar while those in red with $CHF_3$ + $O_2$. (e) $I_D$-$V_{DS}$ characteristics of the device labeled as 1 in **d**, exhibiting almost symmetric and linear (for $V_{DS}$ > 0.5 V) transport behavior up to 3 V. (f) $I_D$-$V_{GS}$ characteristics of identical devices 1 and 2 showing very similar current densities. The dashed curves are fits obtained using the model in Eq. 4 (main text). The respective threshold voltages $V_T$, extracted from the fits, were subtracted from $V_{GS}$ in all plots, for a better comparison. (g) $I_D$-$V_{GS}$ characteristics of device 2 (same as in **f**) plotted on a log-scale, displaying an on/off current ratio >$10^7$ and a steep subthreshold slope (inset), maintained up to three orders of magnitude.

# S4. MoS$_2$ FETs with pure Au edge contacts

In addition to Ti-Au, we also fabricated edge contacts with pure Au (i.e. without any Ti adhesion layer), since Au has been reported to result in low resistance contacts to MoS$_2$[8]. Figure S7 shows the optical images of an hBN encapsulated 1L-MoS$_2$ sample with Au edge contacts, together with the I-V characteristics of one device. However, this sample was among the very few that were successfully fabricated since, in general, Au was found to often result in devices that did not conduct at all. Upon closer inspection of failed samples in an SEM, we noticed gaps between Au and hBN in several contacts, as shown in Fig. S8. By comparing the morphology of Au before and after annealing, we observed that during annealing at 300 °C, Au tends to reflow and lose contact with hBN due to poor adhesion. This occurred despite Au deposition at an angle so as to completely cover the hBN sidewalls. Therefore, we infer that pure Au in combination with high-temperature annealing, is not suitable for edge contacts.

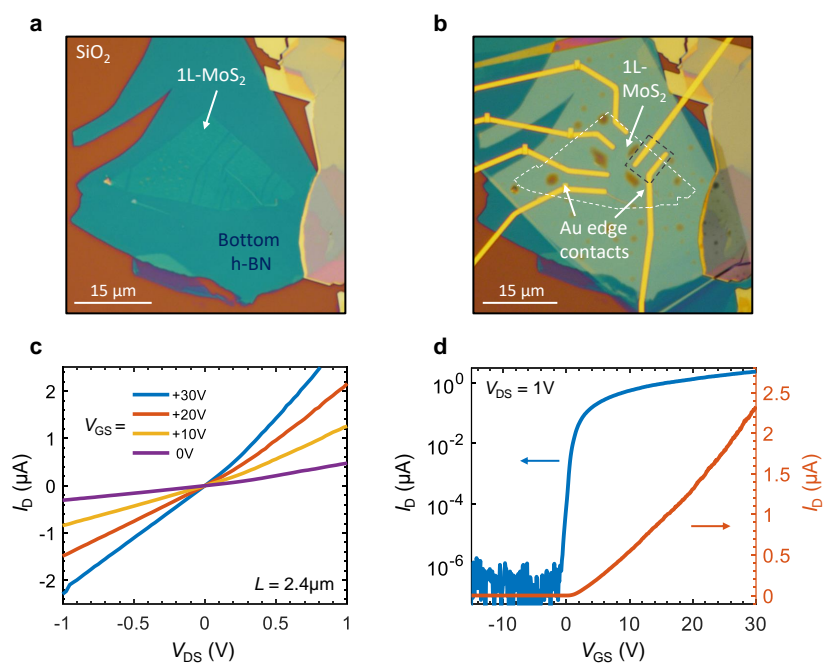

**Figure S7**. **Pure Au edge contacts.** (a) Optical image of a 1L MoS$_2$-hBN heterostructure on a Si/SiO$_2$ (285 nm) substrate. (b) Final stack after top hBN transfer and Au (60 nm) edge contacts fabrication. The encapsulated MoS$_2$ flake has been demarcated by white dashed lines. No Ar$^+$ sputtering was performed for this sample. (c-d) $I_D$-$V_{DS}$ and $I_D$-$V_{GS}$ characteristics of the device outlined by the black dashed rectangle in **b**.

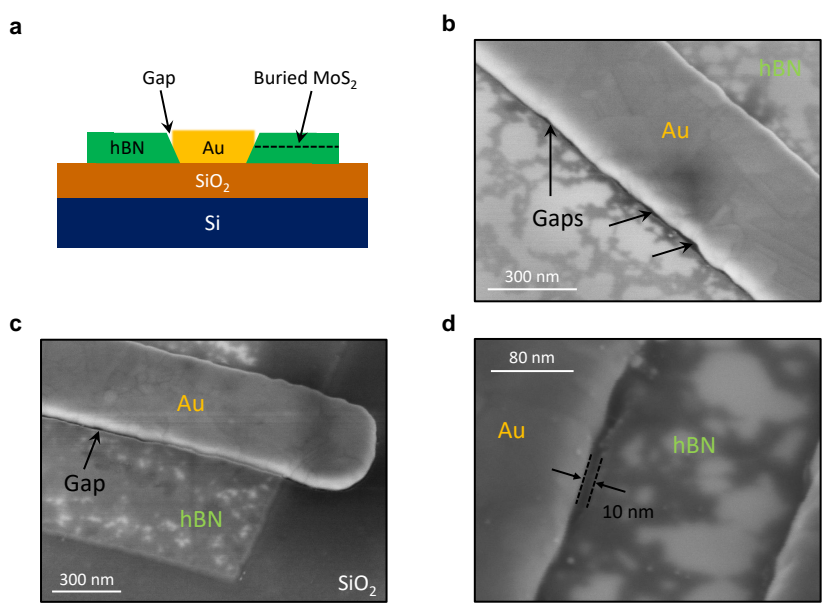

**Figure S8**. **Poor adhesion between Au and hBN.** (a) Schematic illustration of Au losing contact with MoS$_2$ encapsulated between two hBN layers. (b-d) SEM images of actual devices revealing gaps between hBN sidewalls and Au after annealing at 300 °C. Such gaps often cause devices to not conduct, making Au without an adhesion layer, highly unreliable for use in edge contacts.

## S5. Edge contacts without Ar$^+$ sputtering

In the most general case, Ti-MoS$_2$ edge contacts can be considered to be composed of Ti-TiO$_x$-(air molecules)-MoO$_x$-MoS$_2$ junctions. For improving the contact performance, it is important to determine the role of each interfacial species. As discussed in the main text, we fabricated several devices without sputtering the MoS$_2$ contact edges with Ar$^+$, in order to leave air molecules and MoO$_x$ (if any) intact. The transfer characteristics of three such devices, shown in Fig. S9a, exhibit a substantially lower $I_D$ compared to Ar$^+$ sputtered devices fabricated on the same hBN-MoS$_2$-hBN stack (see Fig. 1h of the main text). Furthermore, it has been found previously by X-ray photoelectron spectroscopy (XPS)[9] that Ti can oxidize to form TiO$_x$ when deposited under moderately high vacuum conditions ($\sim 1 \times 10^{-6}$ mbar). To inhibit TiO$_x$ formation and study its effect, Ti-Au was deposited at $\sim 1 \times 10^{-8}$ mbar on another sample (Fig. S9b), again without Ar$^+$ sputtering. Before metal deposition on the contacts, Ti (80 nm) was evaporated into the chamber while keeping the sample surface covered by a shutter, to capture residual O$_2$ molecules and lower the chamber pressure. However, as seen in Fig. S9c, $I_D$ still remains $<1$ μA/μm at $V_{DS} = 1$ V, irrespective of channel length $L$, revealing an $R_C$ dominated transport in all devices. Notably, one device displayed random conductance fluctuations ($L = 1$ μm), implying an unstable contact while two others didn't conduct at all ($L = 0.5, 1.5$ μm). These results underline the importance of Ar$^+$ sputtering in realizing good edge contacts and without it, the deposition pressure (*i.e.* the absence or presence of TiO$_x$), does not make a significant difference.

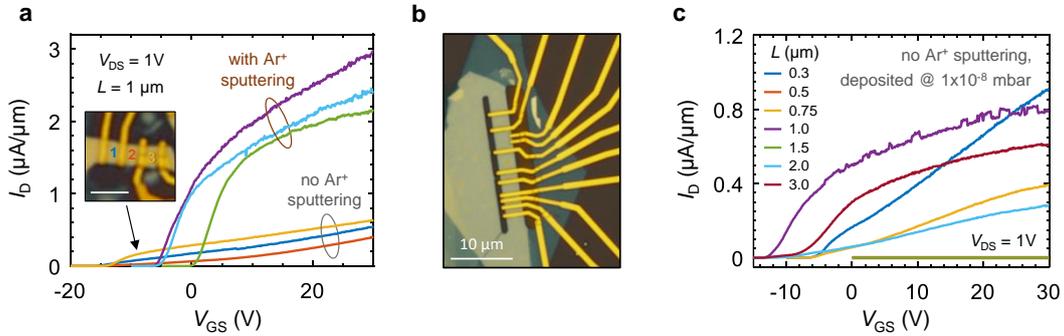

**Figure S9**. **Devices without Ar$^+$ sputtering. (a)** $I_D$-$V_{GS}$ characteristics of three control devices with identical dimensions (inset), also fabricated on the 1L-MoS$_2$ flake presented in Figs. 1-2 of the main text and in Fig. S4, but without performing *in-situ* Ar$^+$ sputtering before metal deposition. The current density in these devices is 5-8x lower compared to those on the same sample that were Ar$^+$ sputtered. Ti-Au deposition was done at $\sim 1 \times 10^{-7}$ mbar. Inset scale bar: 3 μm. **(b)** Optical image of another set of control devices in which, Ar$^+$ sputtering was again not performed and additionally, Ti-Au was deposited at a base pressure of $\sim 1 \times 10^{-8}$ mbar to prevent TiO$_x$ deposition. The channel lengths ($L$) range from 0.25-3 μm. **(c)** $I_D$-$V_{GS}$ characteristics of the devices shown in **b**. Despite the higher vacuum, $I_D$ in this case too is similar to **a** and much lower than that obtained after Ar$^+$ sputtering. Moreover, no correlation with $L$ is discernible, indicating a high $R_C$ in every device. All $I_D$-$V_{GS}$ plots shown in **a** and **c** were recorded after annealing in Ar + H$_2$ at 200 °C for 3 h. The characteristics of un-sputtered devices did not improve considerably upon further annealing at higher temperatures.

## S6. Quantum transport simulations

To gain further insight into carrier transport through edge contacts, we performed ballistic *ab initio* quantum transport simulations based on density functional theory (DFT). The non-equilibrium Green's function (NEGF) formalism was used for this purpose, which involves solving the following equation,

$$[E - H_{\text{MLWF}}(k_z) - \Sigma^{\text{RB}}(E, k_z)] \cdot G^{\text{R}}(E, k_z) = I \quad (1)$$

for the momentum $k_z$ and energy $E$ dependent retarded Green's function $G^{\text{R}}(E, k_z)$ and boundary self-energy $\Sigma^{\text{RB}}(E, k_z)$. The Hamiltonian matrix $H_{\text{MLWF}}(k_z)$ was expressed in a maximally localized Wannier function (MLWF) basis. It was constructed according to the upscaling technique described in Ref. 10 for the Ti-MoS$_2$ edge contact structure depicted schematically in Fig. S10a. The size of $H_{\text{MLWF}}(k_z)$ is equal to $N_A \times N_{\text{orb}}$, where $N_A$ is the number of atoms in the simulation domain and $N_{\text{orb}}$ is the average number of MLWF per atom. Once Eq. 1 was solved for all energies and momenta of interest, the density-of-states $\text{DOS}(E, k_z) = \text{diag}(G^{\text{R}}(E, k_z) - G^{\text{A}}(E, k_z))$ and the transmission function $\mathcal{T}(E, k_z) = \text{tr}(G^{\text{R}}_{1N}(E, k_z) \cdot \Gamma_{NN} \cdot G^{\text{A}}_{N1}(E, k_z) \cdot \Gamma_{11})$ could be computed, from which the charge density and electronic current were derived. Here, $G^{\text{A}}(E, k_z)$ is the advanced Green's function and $\Gamma_{11/NN}$ is the broadening function corresponding to the first (index 1) and last (index $N$) unit cell of the contact geometry.

The resulting spatially- and energetically-resolved $\text{DOS}(E) = \sum_{k_z} \text{DOS}(E, k_z)$ is displayed in Fig. S10b. From this plot, the Ti-MoS$_2$ band alignment and the resulting Schottky barrier height ($\phi_{\text{SB}}$) can be identified. The energy-dependent electron transmission function $\mathcal{T}(E) = \sum_{k_z} \mathcal{T}(E, k_z)$ is plotted in Fig. S10c. The total injected current density was then determined via the Landauer-Büttiker formalism by multiplying $\mathcal{T}(E, k_z)$ with the difference between the Fermi-Dirac distribution functions $f(E, T)$ (*i.e.* the occupation probability at a given temperature) of the right and left contacts and integrating the result over energy and momentum. Figure S10d shows the simulated current density as a function of temperature. Since in real devices the contact metal Fermi level is often strongly pinned close to the MoS$_2$ conduction band minimum[11,12] and because DFT, in general, does not accurately estimate the band offset between two materials, the current was calculated for several values of $\phi_{\text{SB}}$. This was done by artificially raising the metal Fermi level indicated in Fig. S10b. As evident from Fig. S10d, the injected current density remains largely independent of temperature, regardless of the barrier height, which corroborates the measured $I_D$-$V_{\text{GS}}$ characteristics in Fig. 4a of the main text.

The relative temperature insensitivity of the current can be traced back to the underlying carrier injection mechanism and the shape of the transmission function $\mathcal{T}(E)$. In case of Ti-MoS$_2$ edge contacts, the largest contribution to the current does not come from thermionic processes but instead from quantum mechanical tunneling through the Schottky barrier, which does not heavily depend on temperature. Furthermore, the magnitude of the transmission function from Ti to MoS$_2$ does not vary much around the Fermi energy such that the current $I_D$ for large $V_{\text{DS}}$ can be approximated as, $I_D \approx \frac{e}{\hbar} \int_{E_{\text{min}}}^{\infty} C \cdot f(E, E_F^M, T) \, dE$, where $E_{\text{min}}$ corresponds to the conduction band minimum of MoS$_2$ far from the interface region, $E_F^M$ is the Fermi level of the metal contact and $C$ is a constant. It can be shown that such an expression exhibits little to no dependence on the temperature $T$.

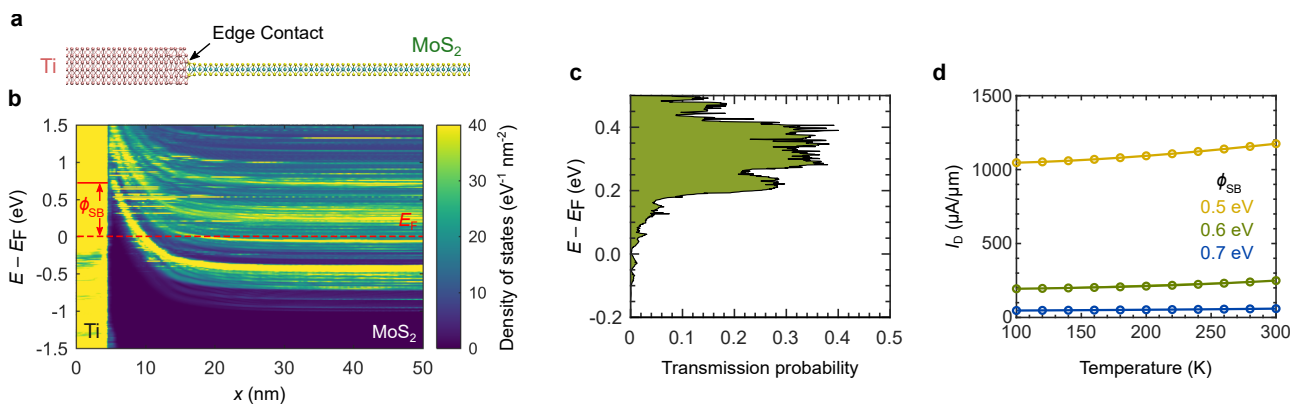

**Figure S10**. **Quantum transport simulations of edge contacts.** (a) Schematic cross-section and (b) computed band-diagram of a Ti - monolayer MoS$_2$ edge contact structure. The color scale in (b) represents the DOS per unit area plotted on a linear scale, the horizontal red dashed line denotes the Fermi level $E_F$ and $\phi_{\text{SB}}$ indicates the Schottky barrier height. The effect of an applied gate bias was incorporated by shifting the MoS$_2$ conduction band minima to 1 eV below $E_F$. (c) Energy dependent electron transmission probability $\mathcal{T}(E)$ for the DOS distribution shown in **b**. (d) Total current density as a function of temperature, calculated for three different $\phi_{\text{SB}}$ values. We find that the injected current density depends only weakly on temperature for T<300 K since the transmission function in **c** does not change appreciably around the Fermi level.




\begin{thebibliography}{10}
	\expandafter\ifx\csname url\endcsname\relax
	\def\url#1{\texttt{#1}}\fi
	\expandafter\ifx\csname urlprefix\endcsname\relax\def\urlprefix{URL }\fi
	\providecommand{\bibinfo}[2]{#2}
	\providecommand{\eprint}[2][]{\url{#2}}
	\providecommand \doibase [0]{http://dx.doi.org/}%
	
	\bibitem{Desai16}
	\bibinfo{author}{Desai, S.~B.} \emph{et~al.}
	\newblock \bibinfo{title}{{MoS}$_{\rm 2}$ transistors with 1-nanometer gate
		lengths}.
	\newblock \href {\doibase 10.1126/science.aah4698} {
		\emph{\bibinfo{journal}{Science}} \textbf{\bibinfo{volume}{354}},
		\bibinfo{pages}{99--103} (\bibinfo{year}{2016})}.
	
	\bibitem{Iannaccone18}
	\bibinfo{author}{Iannaccone, G.}, \bibinfo{author}{Bonaccorso, F.},
	\bibinfo{author}{Colombo, L.} \& \bibinfo{author}{Fiori, G.}
	\newblock \bibinfo{title}{Quantum engineering of transistors based on {2D}
		materials heterostructures}.
	\newblock \href {\doibase 10.1038/s41565-018-0082-6} {
		\emph{\bibinfo{journal}{Nature Nanotech.}} \textbf{\bibinfo{volume}{13}},
		\bibinfo{pages}{183--191} (\bibinfo{year}{2018})}.
	
	\bibitem{Frisenda2018b}
	\bibinfo{author}{Frisenda, R.}, \bibinfo{author}{Molina-Mendoza, A.~J.},
	\bibinfo{author}{Mueller, T.}, \bibinfo{author}{Castellanos-Gomez, A.} \&
	\bibinfo{author}{van~der Zant, H. S.~J.}
	\newblock \bibinfo{title}{Atomically thin p–n junctions based on
		two-dimensional materials}.
	\newblock \href {\doibase 10.1039/C7CS00880E} {
		\emph{\bibinfo{journal}{Chem. Soc. Rev.}} \textbf{\bibinfo{volume}{47}},
		\bibinfo{pages}{3339--3358} (\bibinfo{year}{2018})}.
	
	\bibitem{Wang2018b}
	\bibinfo{author}{Wang, M.} \emph{et~al.}
	\newblock \bibinfo{title}{Robust memristors based on layered two-dimensional
		materials}.
	\newblock \href {\doibase 10.1038/s41928-018-0021-4} {
		\emph{\bibinfo{journal}{Nature Electron.}} \textbf{\bibinfo{volume}{1}},
		\bibinfo{pages}{130--136} (\bibinfo{year}{2018})}.
	
	\bibitem{Bharadwaj15}
	\bibinfo{author}{Bharadwaj, P.} \& \bibinfo{author}{Novotny, L.}
	\newblock \bibinfo{title}{Optoelectronics in flatland}.
	\newblock \href {\doibase 10.1364/OPN.26.7.000024} {
		\emph{\bibinfo{journal}{Optics and Photonics News}}
		\textbf{\bibinfo{volume}{26}}, \bibinfo{pages}{24--31}
		(\bibinfo{year}{2015})}.
	
	\bibitem{Wang2018}
	\bibinfo{author}{Wang, J.}, \bibinfo{author}{Verzhbitskiy, I.} \&
	\bibinfo{author}{Eda, G.}
	\newblock \bibinfo{title}{Electroluminescent devices based on {2D}
		semiconducting transition metal dichalcogenides}.
	\newblock \href {\doibase 10.1002/adma.201802687} {
		\emph{\bibinfo{journal}{Adv. Mater.}} \textbf{\bibinfo{volume}{30}},
		\bibinfo{pages}{1802687} (\bibinfo{year}{2018})}.
	
	\bibitem{Schaibley16}
	\bibinfo{author}{Schaibley, J.~R.} \emph{et~al.}
	\newblock \bibinfo{title}{Valleytronics in {2D} materials}.
	\newblock \href {\doibase 10.1038/natrevmats.2016.55} {
		\emph{\bibinfo{journal}{Nat. Rev. Mater.}} \textbf{\bibinfo{volume}{1}},
		\bibinfo{pages}{16055} (\bibinfo{year}{2016})}.
	
	\bibitem{Qiu12}
	\bibinfo{author}{Qiu, H.} \emph{et~al.}
	\newblock \bibinfo{title}{Electrical characterization of back-gated bi-layer
		{MoS}$_{\rm 2}$ field-effect transistors and the effect of ambient on their
		performances}.
	\newblock \href {\doibase 10.1063/1.3696045} {
		\emph{\bibinfo{journal}{Appl. Phys. Lett.}} \textbf{\bibinfo{volume}{100}},
		\bibinfo{pages}{98--101} (\bibinfo{year}{2012})}.
	
	\bibitem{Jariwala13}
	\bibinfo{author}{Jariwala, D.} \emph{et~al.}
	\newblock \bibinfo{title}{Band-like transport in high mobility unencapsulated
		single-layer {MoS}$_{\rm 2}$ transistors}.
	\newblock \href {\doibase 10.1063/1.4803920} {
		\emph{\bibinfo{journal}{Appl. Phys. Lett.}} \textbf{\bibinfo{volume}{102}},
		\bibinfo{pages}{4--8} (\bibinfo{year}{2013})}.
	
	\bibitem{Park13}
	\bibinfo{author}{Park, W.} \emph{et~al.}
	\newblock \bibinfo{title}{Oxygen environmental and passivation effects on
		molybdenum disulfide field effect transistors}.
	\newblock \href {\doibase 10.1088/0957-4484/24/9/095202} {
		\emph{\bibinfo{journal}{Nanotechnology}} \textbf{\bibinfo{volume}{24}},
		\bibinfo{pages}{095202} (\bibinfo{year}{2013})}.
	
	\bibitem{Wu16}
	\bibinfo{author}{Wu, D.} \emph{et~al.}
	\newblock \bibinfo{title}{Uncovering edge states and electrical inhomogeneity
		in {MoS}$_{\rm 2}$ field-effect transistors}.
	\newblock \href {\doibase 10.1073/pnas.1605982113} {
		\emph{\bibinfo{journal}{Proc. Natl. Acad. Sci. USA}}
		\textbf{\bibinfo{volume}{113}}, \bibinfo{pages}{8583--8588}
		(\bibinfo{year}{2016})}.
	
	\bibitem{Rahimi16}
	\bibinfo{author}{Rahimi, S.} \emph{et~al.}
	\newblock \bibinfo{title}{The positive effects of hydrophobic fluoropolymers on
		the electrical properties of {MoS}$_{\rm 2}$ transistors}.
	\newblock \href {\doibase 10.3390/app6090236} {
		\emph{\bibinfo{journal}{Appl. Sci.}} \textbf{\bibinfo{volume}{6}},
		\bibinfo{pages}{236} (\bibinfo{year}{2016})}.
	
	\bibitem{Smithe2017b}
	\bibinfo{author}{Smithe, K. K.~H.}, \bibinfo{author}{Suryavanshi, S.~V.},
	\bibinfo{author}{Mu{\~n}oz~Rojo, M.}, \bibinfo{author}{Tedjarati, A.~D.} \&
	\bibinfo{author}{Pop, E.}
	\newblock \bibinfo{title}{Low variability in synthetic monolayer {MoS}$_{\rm
			2}$ devices}.
	\newblock \href {\doibase 10.1021/acsnano.7b04100} {
		\emph{\bibinfo{journal}{ACS Nano}} \textbf{\bibinfo{volume}{11}},
		\bibinfo{pages}{8456--8463} (\bibinfo{year}{2017})}.
	
	\bibitem{Giannazzo15}
	\bibinfo{author}{Giannazzo, F.}, \bibinfo{author}{Fisichella, G.},
	\bibinfo{author}{Piazza, A.}, \bibinfo{author}{Agnello, S.} \&
	\bibinfo{author}{Roccaforte, F.}
	\newblock \bibinfo{title}{Nanoscale inhomogeneity of the {S}chottky barrier and
		resistivity in {MoS}$_{\rm 2}$ multilayers}.
	\newblock \href {\doibase 10.1103/PhysRevB.92.081307} {
		\emph{\bibinfo{journal}{Phys. Rev. B}} \textbf{\bibinfo{volume}{92}},
		\bibinfo{pages}{081307} (\bibinfo{year}{2015})}.
	
	\bibitem{Ji2018}
	\bibinfo{author}{Ji, H.} \emph{et~al.}
	\newblock \bibinfo{title}{Gas adsorbates are coulomb scatterers{,} rather than
		neutral ones{,} in a monolayer {MoS}$_{\rm 2}$ field effect transistor}.
	\newblock \href {\doibase 10.1039/C8NR03570A} {
		\emph{\bibinfo{journal}{Nanoscale}} \textbf{\bibinfo{volume}{10}},
		\bibinfo{pages}{10856--10862} (\bibinfo{year}{2018})}.
	
	\bibitem{Sangwan13}
	\bibinfo{author}{Sangwan, V.~K.} \emph{et~al.}
	\newblock \bibinfo{title}{Low-frequency electronic noise in single-layer
		{MoS}$_{\rm 2}$ transistors}.
	\newblock \href {\doibase 10.1021/nl402150r} {
		\emph{\bibinfo{journal}{Nano Letters}} \textbf{\bibinfo{volume}{13}},
		\bibinfo{pages}{4351--4355} (\bibinfo{year}{2013})}.
	
	\bibitem{Xie14}
	\bibinfo{author}{Xie, X.} \emph{et~al.}
	\newblock \bibinfo{title}{Low-frequency noise in bilayer {MoS}$_{\rm 2}$
		transistor}.
	\newblock \href {\doibase 10.1021/nn4066473} {
		\emph{\bibinfo{journal}{ACS Nano}} \textbf{\bibinfo{volume}{8}},
		\bibinfo{pages}{5633--5640} (\bibinfo{year}{2014})}.
	
	\bibitem{Shimazu16}
	\bibinfo{author}{Shimazu, Y.}, \bibinfo{author}{Tashiro, M.},
	\bibinfo{author}{Sonobe, S.} \& \bibinfo{author}{Takahashi, M.}
	\newblock \bibinfo{title}{Environmental effects on hysteresis of transfer
		characteristics in molybdenum disulfide field-effect transistors}.
	\newblock \href {\doibase 10.1038/srep30084} {
		\emph{\bibinfo{journal}{Sci. Rep.}} \textbf{\bibinfo{volume}{6}},
		\bibinfo{pages}{6--11} (\bibinfo{year}{2016})}.
	
	\bibitem{Cho14}
	\bibinfo{author}{Cho, K.} \emph{et~al.}
	\newblock \bibinfo{title}{Gate-bias stress-dependent photoconductive
		characteristics of multi-layer {MoS}$_{\rm 2}$ field-effect transistors}.
	\newblock \href {\doibase 10.1088/0957-4484/25/15/155201} {
		\emph{\bibinfo{journal}{Nanotechnology}} \textbf{\bibinfo{volume}{25}},
		\bibinfo{pages}{155201} (\bibinfo{year}{2014})}.
	
	\bibitem{Baugher13}
	\bibinfo{author}{Baugher, B.}, \bibinfo{author}{Churchill, H. O.~H.},
	\bibinfo{author}{Yang, Y.} \& \bibinfo{author}{Jarillo-Herrero, P.}
	\newblock \bibinfo{title}{Intrinsic electronic transport properties of high
		quality monolayer and bilayer {MoS}$_{\rm 2}$}.
	\newblock \href {\doibase 10.1021/nl401916s} {
		\emph{\bibinfo{journal}{Nano Lett.}} \textbf{\bibinfo{volume}{13}},
		\bibinfo{pages}{4212--4216} (\bibinfo{year}{2013})}.
	
	\bibitem{Smithe2017a}
	\bibinfo{author}{Smithe, K.~K.}, \bibinfo{author}{English, C.~D.},
	\bibinfo{author}{Suryavanshi, S.~V.} \& \bibinfo{author}{Pop, E.}
	\newblock \bibinfo{title}{Intrinsic electrical transport and performance
		projections of synthetic monolayer {MoS}$_{\rm 2}$ devices}.
	\newblock \href {\doibase 10.1088/2053-1583/4/1/011009} {
		\emph{\bibinfo{journal}{2D Materials}} \textbf{\bibinfo{volume}{4}},
		\bibinfo{pages}{1--8} (\bibinfo{year}{2017})}.
	
	\bibitem{Peto2018}
	\bibinfo{author}{Peto, J.} \emph{et~al.}
	\newblock \bibinfo{title}{Spontaneous doping of the basal plane of {MoS}$_{\rm
			2}$ single layers through oxygen substitution under ambient conditions}.
	\newblock \href {\doibase 10.1038/s41557-018-0136-2} {
		\emph{\bibinfo{journal}{Nature Chem.}} \textbf{\bibinfo{volume}{10}},
		\bibinfo{pages}{1246--1251} (\bibinfo{year}{2018})}.
	
	\bibitem{Lee15}
	\bibinfo{author}{Lee, G.-H.} \emph{et~al.}
	\newblock \bibinfo{title}{Highly stable, dual-gated {MoS}$_{\rm 2}$ transistors
		encapsulated by hexagonal boron nitride with gate-controllable contact
		resistance and threshold voltage}.
	\newblock \href {\doibase 10.1021/acsnano.5b01341} {
		\emph{\bibinfo{journal}{ACS Nano}} \textbf{\bibinfo{volume}{9}},
		\bibinfo{pages}{7019--7026} (\bibinfo{year}{2015})}.
	
	\bibitem{Park16}
	\bibinfo{author}{Park, J.~H.} \emph{et~al.}
	\newblock \bibinfo{title}{Scanning tunneling microscopy and spectroscopy of air
		exposure effects on molecular beam epitaxy grown {WSe}$_{\rm 2}$ monolayers
		and bilayers}.
	\newblock \href {\doibase 10.1021/acsnano.5b07698} {
		\emph{\bibinfo{journal}{ACS Nano}} \textbf{\bibinfo{volume}{10}},
		\bibinfo{pages}{4258--4267} (\bibinfo{year}{2016})}.
	
	\bibitem{Gioele16}
	\bibinfo{author}{Mirabelli, G.} \emph{et~al.}
	\newblock \bibinfo{title}{Air sensitivity of {MoS}$_{\rm 2}$, {MoSe}$_{\rm 2}$,
		{MoTe}$_{\rm 2}$, {HfS}$_{\rm 2}$, and {HfSe}$_{\rm 2}$}.
	\newblock \href {\doibase 10.1063/1.4963290} {
		\emph{\bibinfo{journal}{J. Appl. Phys.}} \textbf{\bibinfo{volume}{120}},
		\bibinfo{pages}{125102} (\bibinfo{year}{2016})}.
	
	\bibitem{Gao16}
	\bibinfo{author}{Gao, J.} \emph{et~al.}
	\newblock \bibinfo{title}{Aging of transition metal dichalcogenide monolayers}.
	\newblock \href {\doibase 10.1021/acsnano.5b07677} {
		\emph{\bibinfo{journal}{ACS Nano}} \textbf{\bibinfo{volume}{10}},
		\bibinfo{pages}{2628--2635} (\bibinfo{year}{2016})}.
	
	\bibitem{Cao15}
	\bibinfo{author}{Cao, Y.} \emph{et~al.}
	\newblock \bibinfo{title}{Quality heterostructures from two-dimensional
		crystals unstable in air by their assembly in inert atmosphere}.
	\newblock \href {\doibase 10.1021/acs.nanolett.5b00648} {
		\emph{\bibinfo{journal}{Nano Lett.}} \textbf{\bibinfo{volume}{15}},
		\bibinfo{pages}{4914--4921} (\bibinfo{year}{2015})}.
	
	\bibitem{Cui15}
	\bibinfo{author}{Cui, X.} \emph{et~al.}
	\newblock \bibinfo{title}{Multi-terminal transport measurements of {MoS}$_{\rm
			2}$ using a van der {W}aals heterostructure device platform}.
	\newblock \href {\doibase 10.1038/nnano.2015.70} {
		\emph{\bibinfo{journal}{Nature Nanotech.}} \textbf{\bibinfo{volume}{10}},
		\bibinfo{pages}{534--540} (\bibinfo{year}{2015})}.
	
	\bibitem{Liu2015a}
	\bibinfo{author}{Liu, Y.} \emph{et~al.}
	\newblock \bibinfo{title}{Toward barrier free contact to molybdenum disulfide
		using graphene electrodes}.
	\newblock \href {\doibase 10.1021/nl504957p} {
		\emph{\bibinfo{journal}{Nano Lett.}} \textbf{\bibinfo{volume}{15}},
		\bibinfo{pages}{3030--3034} (\bibinfo{year}{2015})}.
	
	\bibitem{Guan17}
	\bibinfo{author}{Guan, J.}, \bibinfo{author}{Chuang, H.~J.},
	\bibinfo{author}{Zhou, Z.} \& \bibinfo{author}{Tom{\'{a}}nek, D.}
	\newblock \bibinfo{title}{Optimizing charge injection across transition metal
		dichalcogenide heterojunctions: Theory and experiment}.
	\newblock \href {\doibase 10.1021/acsnano.7b00285} {
		\emph{\bibinfo{journal}{ACS Nano}} \textbf{\bibinfo{volume}{11}},
		\bibinfo{pages}{3904--3910} (\bibinfo{year}{2017})}.
	
	\bibitem{Sata17}
	\bibinfo{author}{Sata, Y.} \emph{et~al.}
	\newblock \bibinfo{title}{N- and p-type carrier injections into {WSe}$_{\rm 2}$
		with van der {W}aals contacts of two-dimensional materials}.
	\newblock \href {\doibase 10.7567/JJAP.56.04CK09} {
		\emph{\bibinfo{journal}{Jpn. J. Appl. Phys.}} \textbf{\bibinfo{volume}{56}},
		\bibinfo{pages}{04CK09} (\bibinfo{year}{2017})}.
	
	\bibitem{Wang15b}
	\bibinfo{author}{Wang, J. I.-J.} \emph{et~al.}
	\newblock \bibinfo{title}{Electronic transport of encapsulated graphene and
		{WSe}$_{\rm 2}$ devices fabricated by pick-up of prepatterned {hBN}}.
	\newblock \href {\doibase 10.1021/nl504750f} {
		\emph{\bibinfo{journal}{Nano Lett.}} \textbf{\bibinfo{volume}{15}},
		\bibinfo{pages}{1898--1903} (\bibinfo{year}{2015})}.
	
	\bibitem{Telford18}
	\bibinfo{author}{Telford, E.~J.} \emph{et~al.}
	\newblock \bibinfo{title}{Via method for lithography free contact and
		preservation of {2D} materials}.
	\newblock \href {\doibase 10.1021/acs.nanolett.7b05161} {
		\emph{\bibinfo{journal}{Nano Lett.}} \textbf{\bibinfo{volume}{18}},
		\bibinfo{pages}{1416--1420} (\bibinfo{year}{2018})}.
	
	\bibitem{Liu2018}
	\bibinfo{author}{Liu, Y.} \emph{et~al.}
	\newblock \bibinfo{title}{Approaching the {S}chottky-{M}ott limit in van der
		{W}aals metal-semiconductor junctions}.
	\newblock \href {\doibase 10.1038/s41586-018-0129-8} {
		\emph{\bibinfo{journal}{Nature}} \textbf{\bibinfo{volume}{557}},
		\bibinfo{pages}{696--700} (\bibinfo{year}{2018})}.
	
	\bibitem{Liao2018}
	\bibinfo{author}{Liao, M.} \emph{et~al.}
	\newblock \bibinfo{title}{Twist angle-dependent conductivities across
		{MoS}$_{\rm 2}$/graphene heterojunctions}.
	\newblock \href {\doibase 10.1038/s41467-018-06555-w} {
		\emph{\bibinfo{journal}{Nature Commun.}} \textbf{\bibinfo{volume}{9}},
		\bibinfo{pages}{4068} (\bibinfo{year}{2018})}.
	
	\bibitem{Ling16}
	\bibinfo{author}{Ling, X.} \emph{et~al.}
	\newblock \bibinfo{title}{Parallel stitching of {2D} materials}.
	\newblock \href {\doibase 10.1002/adma.201505070} {
		\emph{\bibinfo{journal}{Adv. Mater.}} \textbf{\bibinfo{volume}{28}},
		\bibinfo{pages}{2322--2329} (\bibinfo{year}{2016})}.
	
	\bibitem{Guimaraes16}
	\bibinfo{author}{Guimarães, M.~H.} \emph{et~al.}
	\newblock \bibinfo{title}{Atomically thin {O}hmic edge contacts between
		two-dimensional materials}.
	\newblock \href {\doibase 10.1021/acsnano.6b02879} {
		\emph{\bibinfo{journal}{ACS Nano}} \textbf{\bibinfo{volume}{10}},
		\bibinfo{pages}{6392--6399} (\bibinfo{year}{2016})}.
	
	\bibitem{Zhao16}
	\bibinfo{author}{Zhao, M.} \emph{et~al.}
	\newblock \bibinfo{title}{Large-scale chemical assembly of atomically thin
		transistors and circuits}.
	\newblock \href {\doibase 10.1038/nnano.2016.115} {
		\emph{\bibinfo{journal}{Nature Nanotech.}} \textbf{\bibinfo{volume}{11}},
		\bibinfo{pages}{954--959} (\bibinfo{year}{2016})}.
	
	\bibitem{Suenaga2018}
	\bibinfo{author}{Suenaga, K.} \emph{et~al.}
	\newblock \bibinfo{title}{Surface-mediated aligned growth of monolayer
		{MoS}$_{\rm 2}$ and in-plane heterostructures with graphene on sapphire}.
	\newblock \href {\doibase 10.1021/acsnano.8b04612} {
		\emph{\bibinfo{journal}{ACS Nano}} \textbf{\bibinfo{volume}{12}},
		\bibinfo{pages}{10032--10044} (\bibinfo{year}{2018})}.
	
	\bibitem{Han2018}
	\bibinfo{author}{Han, Y.} \emph{et~al.}
	\newblock \bibinfo{title}{Strain mapping of two-dimensional heterostructures
		with subpicometer precision}.
	\newblock \href {\doibase 10.1021/acs.nanolett.8b00952} {
		\emph{\bibinfo{journal}{Nano Lett.}} \textbf{\bibinfo{volume}{18}},
		\bibinfo{pages}{3746--3751} (\bibinfo{year}{2018})}.
	
	\bibitem{Wang16}
	\bibinfo{author}{Wang, J.} \emph{et~al.}
	\newblock \bibinfo{title}{High mobility {MoS}$_{\rm 2}$ transistor with low
		{S}chottky barrier contact by using atomic thick h-{BN} as a tunneling
		layer}.
	\newblock \href {\doibase 10.1002/adma.201602757} {
		\emph{\bibinfo{journal}{Adv. Mater.}} \textbf{\bibinfo{volume}{28}},
		\bibinfo{pages}{8302--8308} (\bibinfo{year}{2016})}.
	
	\bibitem{Cui17}
	\bibinfo{author}{Cui, X.} \emph{et~al.}
	\newblock \bibinfo{title}{Low temperature {O}hmic contact to monolayer
		{MoS}$_{\rm 2}$ by van der {W}aals bonded {Co/h-BN} electrodes}.
	\newblock \href {\doibase 10.1021/acs.nanolett.7b01536} {
		\emph{\bibinfo{journal}{Nano Lett.}} \textbf{\bibinfo{volume}{17}},
		\bibinfo{pages}{4781--4786} (\bibinfo{year}{2017})}.
	
	\bibitem{Li17}
	\bibinfo{author}{Li, X.~X.} \emph{et~al.}
	\newblock \bibinfo{title}{Gate-controlled reversible rectifying behaviour in
		tunnel contacted atomically-thin {MoS}$_{\rm 2}$ transistor}.
	\newblock \href {\doibase 10.1038/s41467-017-01128-9} {
		\emph{\bibinfo{journal}{Nature Commun.}} \textbf{\bibinfo{volume}{8}}
		(\bibinfo{year}{2017})}.
	
	\bibitem{Ghiasi2018}
	\bibinfo{author}{Ghiasi, T.~S.}, \bibinfo{author}{Quereda, J.} \&
	\bibinfo{author}{van Wees, B.~J.}
	\newblock \bibinfo{title}{Bilayer h-{BN} barriers for tunneling contacts in
		fully-encapsulated monolayer {MoSe}$_{\rm 2}$ field-effect transistors}.
	\newblock \href {\doibase 10.1088/2053-1583/aadf47} {
		\emph{\bibinfo{journal}{2D Materials}} \textbf{\bibinfo{volume}{6}},
		\bibinfo{pages}{015002} (\bibinfo{year}{2018})}.
	
	\bibitem{Lee2016}
	\bibinfo{author}{Lee, S.}, \bibinfo{author}{Tang, A.}, \bibinfo{author}{Aloni,
		S.} \& \bibinfo{author}{Philip~Wong, H.-S.}
	\newblock \bibinfo{title}{Statistical study on the {S}chottky barrier reduction
		of tunneling contacts to {CVD} synthesized {MoS}$_{\rm 2}$}.
	\newblock \href {\doibase 10.1021/acs.nanolett.5b03727} {
		\emph{\bibinfo{journal}{Nano Lett.}} \textbf{\bibinfo{volume}{16}},
		\bibinfo{pages}{276--281} (\bibinfo{year}{2016})}.
	
	\bibitem{Wang13}
	\bibinfo{author}{Wang, L.} \emph{et~al.}
	\newblock \bibinfo{title}{One-dimensional electrical contact to a
		two-dimensional material}.
	\newblock \href {\doibase 10.1126/science.1244358} {
		\emph{\bibinfo{journal}{Science}} \textbf{\bibinfo{volume}{342}},
		\bibinfo{pages}{614--617} (\bibinfo{year}{2013})}.
	
	\bibitem{Karpiak2017}
	\bibinfo{author}{Karpiak, B.} \emph{et~al.}
	\newblock \bibinfo{title}{{1D} ferromagnetic edge contacts to {2D}
		graphene/h-{BN} heterostructures}.
	\newblock \href {\doibase 10.1088/2053-1583/aa8d2b} {
		\emph{\bibinfo{journal}{2D Materials}} \textbf{\bibinfo{volume}{5}},
		\bibinfo{pages}{014001} (\bibinfo{year}{2017})}.
	
	\bibitem{Chai16}
	\bibinfo{author}{Chai, Y.} \emph{et~al.}
	\newblock \bibinfo{title}{Making one-dimensional electrical contacts to
		molybdenum disulfide-based heterostructures through plasma etching}.
	\newblock \href {\doibase 10.1002/pssa.201532799} {
		\emph{\bibinfo{journal}{Phys. Stat. Sol. A}} \textbf{\bibinfo{volume}{213}},
		\bibinfo{pages}{1358--1364} (\bibinfo{year}{2016})}.
	
	\bibitem{Moon17}
	\bibinfo{author}{Moon, B.~H.} \emph{et~al.}
	\newblock \bibinfo{title}{Junction-structure-dependent {S}chottky barrier
		inhomogeneity and device ideality of monolayer {MoS}$_{\rm 2}$ field-effect
		transistors}.
	\newblock \href {\doibase 10.1021/acsami.6b16692} {
		\emph{\bibinfo{journal}{ACS Appl. Mater. Interfaces}}
		\textbf{\bibinfo{volume}{9}}, \bibinfo{pages}{11240--11246}
		(\bibinfo{year}{2017})}.
	
	\bibitem{Xu16}
	\bibinfo{author}{Xu, S.} \emph{et~al.}
	\newblock \bibinfo{title}{Universal low-temperature {O}hmic contacts for
		quantum transport in transition metal dichalcogenides}.
	\newblock \href {\doibase 10.1088/2053-1583/3/2/021007} {
		\emph{\bibinfo{journal}{2D Materials}} \textbf{\bibinfo{volume}{3}},
		\bibinfo{pages}{021007} (\bibinfo{year}{2016})}.
	
	\bibitem{Jain2018}
	\bibinfo{author}{Jain, A.} \emph{et~al.}
	\newblock \bibinfo{title}{Minimizing residues and strain in {2D} materials
		transferred from {PDMS}}.
	\newblock \href {\doibase 10.1088/1361-6528/aabd90} {
		\emph{\bibinfo{journal}{Nanotechnology}} \textbf{\bibinfo{volume}{29}},
		\bibinfo{pages}{265203} (\bibinfo{year}{2018})}.
	
	\bibitem{Martincova2017}
	\bibinfo{author}{Martincová, J.}, \bibinfo{author}{Otyepka, M.} \&
	\bibinfo{author}{Lazar, P.}
	\newblock \bibinfo{title}{Is single layer {MoS}$_{\rm 2}$ stable in the air?}
	\newblock \href {\doibase 10.1002/chem.201702860} {
		\emph{\bibinfo{journal}{Chem. Eur. J}} \textbf{\bibinfo{volume}{23}},
		\bibinfo{pages}{13233--13239} (\bibinfo{year}{2017})}.
	
	\bibitem{Santosh2016}
	\bibinfo{author}{K.~C., S.}, \bibinfo{author}{Longo, R.~C.},
	\bibinfo{author}{Addou, R.}, \bibinfo{author}{Wallace, R.~M.} \&
	\bibinfo{author}{Cho, K.}
	\newblock \bibinfo{title}{Electronic properties of {MoS}$_{\rm
			2}$/{MoO}$_{\text{x}}$ interfaces: Implications in tunnel field effect
		transistors and hole contacts}.
	\newblock \href {\doibase 10.1038/srep33562} {
		\emph{\bibinfo{journal}{Sci. Rep.}} \textbf{\bibinfo{volume}{6}},
		\bibinfo{pages}{33562} (\bibinfo{year}{2016})}.
	
	\bibitem{Liu2015b}
	\bibinfo{author}{Liu, W.}, \bibinfo{author}{Sarkar, D.}, \bibinfo{author}{Kang,
		J.}, \bibinfo{author}{Cao, W.} \& \bibinfo{author}{Banerjee, K.}
	\newblock \bibinfo{title}{Impact of contact on the operation and performance of
		back-gated monolayer {MoS}$_{\rm 2}$ field-effect-transistors}.
	\newblock \href {\doibase 10.1021/nn506512j} {
		\emph{\bibinfo{journal}{ACS Nano}} \textbf{\bibinfo{volume}{9}},
		\bibinfo{pages}{7904--7912} (\bibinfo{year}{2015})}.
	
	\bibitem{Guo2015}
	\bibinfo{author}{Guo, Y.}, \bibinfo{author}{Liu, D.} \&
	\bibinfo{author}{Robertson, J.}
	\newblock \bibinfo{title}{{3D} behavior of {S}chottky barriers of {2D}
		transition-metal dichalcogenides}.
	\newblock \href {\doibase 10.1021/acsami.5b06897} {
		\emph{\bibinfo{journal}{ACS Appl. Mater. Interfaces}}
		\textbf{\bibinfo{volume}{7}}, \bibinfo{pages}{25709--25715}
		(\bibinfo{year}{2015})}.
	
	\bibitem{Dong2019}
	\bibinfo{author}{Dong, W.} \& \bibinfo{author}{Littlewood, P.~B.}
	\newblock \bibinfo{title}{Quantum electron transport in {O}hmic edge contacts
		between two-dimensional materials}.
	\newblock \href {\doibase 10.1021/acsaelm.9b00095} {
		\emph{\bibinfo{journal}{ACS Appl. Electron. Mater.}}
		\textbf{\bibinfo{volume}{1}}, \bibinfo{pages}{799--803}
		(\bibinfo{year}{2019})}.
	
	\bibitem{Dean10}
	\bibinfo{author}{Dean, C.~R.} \emph{et~al.}
	\newblock \bibinfo{title}{Boron nitride substrates for high-quality graphene
		electronics}.
	\newblock \href {\doibase 10.1038/nnano.2010.172} {
		\emph{\bibinfo{journal}{Nature Nanotech.}} \textbf{\bibinfo{volume}{5}},
		\bibinfo{pages}{722--726} (\bibinfo{year}{2010})}.
	
	\bibitem{Sze2006}
	\bibinfo{author}{Sze, S.~M.} \& \bibinfo{author}{Ng, K.~K.}
	\newblock \emph{\bibinfo{title}{Physics of Semiconductor Devices, 3rd Edition}}
	(\bibinfo{publisher}{John Wiley \& Sons}, \bibinfo{year}{2006}).
	
	\bibitem{Zou2014}
	\bibinfo{author}{Zou, X.} \emph{et~al.}
	\newblock \bibinfo{title}{Interface engineering for high-performance top-gated
		{MoS}$_{\rm 2}$ field-effect transistors}.
	\newblock \href {\doibase 10.1002/adma.201402008} {
		\emph{\bibinfo{journal}{Adv. Mater.}} \textbf{\bibinfo{volume}{26}},
		\bibinfo{pages}{6255--6261} (\bibinfo{year}{2014})}.
	
	\bibitem{Choi2015}
	\bibinfo{author}{Choi, K.} \emph{et~al.}
	\newblock \bibinfo{title}{Trap density probing on top-gate {MoS}$_{\rm 2}$
		nanosheet field-effect transistors by photo-excited charge collection
		spectroscopy}.
	\newblock \href {\doibase 10.1039/C4NR06707J} {
		\emph{\bibinfo{journal}{Nanoscale}} \textbf{\bibinfo{volume}{7}},
		\bibinfo{pages}{5617--5623} (\bibinfo{year}{2015})}.
	
	\bibitem{McDonnell16}
	\bibinfo{author}{McDonnell, S.}, \bibinfo{author}{Smyth, C.},
	\bibinfo{author}{Hinkle, C.~L.} \& \bibinfo{author}{Wallace, R.~M.}
	\newblock \bibinfo{title}{{MoS}$_{2}$-titanium contact interface reactions}.
	\newblock \href {\doibase 10.1021/acsami.6b00275} {
		\emph{\bibinfo{journal}{ACS Appl. Mater. Interfaces}}
		\textbf{\bibinfo{volume}{8}}, \bibinfo{pages}{8289--8294}
		(\bibinfo{year}{2016})}.
	
	\bibitem{Ghibaudo1988}
	\bibinfo{author}{Ghibaudo, G.}
	\newblock \bibinfo{title}{New method for the extraction of {MOSFET}
		parameters}.
	\newblock \href {\doibase 10.1049/el:19880369} {
		\emph{\bibinfo{journal}{Electron. Lett.}} \textbf{\bibinfo{volume}{24}},
		\bibinfo{pages}{543--545} (\bibinfo{year}{1988})}.
	
	\bibitem{Jain1988}
	\bibinfo{author}{Jain, S.}
	\newblock \bibinfo{title}{{Measurement of threshold voltage and channel length
			of submicron MOSFETs}}.
	\newblock \href {\doibase 10.1049/ip-i-1.1988.0029} {
		\emph{\bibinfo{journal}{IEE Proc. I - Solid-State Electron Devices}}
		\textbf{\bibinfo{volume}{135}}, \bibinfo{pages}{162--164}
		(\bibinfo{year}{1988})}.
	
	\bibitem{Chang2014}
	\bibinfo{author}{Chang, H.~Y.}, \bibinfo{author}{Zhu, W.} \&
	\bibinfo{author}{Akinwande, D.}
	\newblock \bibinfo{title}{{On the mobility and contact resistance evaluation
			for transistors based on {MoS}$_{2}$ or two-dimensional semiconducting atomic
			crystals}}.
	\newblock \href {\doibase 10.1063/1.4868536} {
		\emph{\bibinfo{journal}{Appl. Phys. Lett.}} \textbf{\bibinfo{volume}{104}}
		(\bibinfo{year}{2014})}.
	
	\bibitem{Cho2018}
	\bibinfo{author}{Cho, K.} \emph{et~al.}
	\newblock \bibinfo{title}{Contact-engineered electrical properties of
		{MoS}$_{\rm 2}$ field-effect transistors via selectively deposited
		thiol-molecules}.
	\newblock \href {\doibase 10.1002/adma.201705540} {
		\emph{\bibinfo{journal}{Adv. Mater.}} \textbf{\bibinfo{volume}{30}},
		\bibinfo{pages}{1705540} (\bibinfo{year}{2018})}.
	
	\bibitem{Szabo2019}
	\bibinfo{author}{Szabó, A.}, \bibinfo{author}{Jain, A.},
	\bibinfo{author}{Parzefall, M.}, \bibinfo{author}{Novotny, L.} \&
	\bibinfo{author}{Luisier, M.}
	\newblock \bibinfo{title}{Electron transport through metal/{MoS}$_{\rm 2}$
		interfaces: Edge- or area-dependent process?}
	\newblock \href {\doibase 10.1021/acs.nanolett.9b00678} {
		\emph{\bibinfo{journal}{Nano Lett.}} \textbf{\bibinfo{volume}{19}},
		\bibinfo{pages}{3641--3647} (\bibinfo{year}{2019})}.
	
	\bibitem{Liu2014c}
	\bibinfo{author}{Liu, H.} \emph{et~al.}
	\newblock \bibinfo{title}{Switching mechanism in single-layer molybdenum
		disulfide transistors: An insight into current flow across {S}chottky
		barriers}.
	\newblock \href {\doibase 10.1021/nn405916t} {
		\emph{\bibinfo{journal}{ACS Nano}} \textbf{\bibinfo{volume}{8}},
		\bibinfo{pages}{1031--1038} (\bibinfo{year}{2014})}.
	
	\bibitem{Allain2015}
	\bibinfo{author}{Allain, A.}, \bibinfo{author}{Kang, J.},
	\bibinfo{author}{Banerjee, K.} \& \bibinfo{author}{Kis, A.}
	\newblock \bibinfo{title}{Electrical contacts to two-dimensional
		semiconductors}.
	\newblock \href {\doibase 10.1038/nmat4452} {
		\emph{\bibinfo{journal}{Nature Mat.}} \textbf{\bibinfo{volume}{14}},
		\bibinfo{pages}{1195--1205} (\bibinfo{year}{2015})}.
	
	\bibitem{Autore2018}
	\bibinfo{author}{Autore, M.} \emph{et~al.}
	\newblock \bibinfo{title}{Boron nitride nanoresonators for phonon-enhanced
		molecular vibrational spectroscopy at the strong coupling limit}.
	\newblock \href {\doibase 10.1038/lsa.2017.172} {
		\emph{\bibinfo{journal}{Light Sci. Appl.}} \textbf{\bibinfo{volume}{7}},
		\bibinfo{pages}{17172} (\bibinfo{year}{2018})}.
	
	\bibitem{Bollinger2001}
	\bibinfo{author}{Bollinger, M.~V.} \emph{et~al.}
	\newblock \bibinfo{title}{One-dimensional metallic edge states in {MoS}$_{\rm
			2}$}.
	\newblock \href {\doibase 10.1103/PhysRevLett.87.196803} {
		\emph{\bibinfo{journal}{Phys. Rev. Lett.}} \textbf{\bibinfo{volume}{87}},
		\bibinfo{pages}{196803} (\bibinfo{year}{2001})}.
	
	\bibitem{Addou2018}
	\bibinfo{author}{Addou, R.} \emph{et~al.}
	\newblock \bibinfo{title}{One dimensional metallic edges in atomically thin
		{WSe}$_{\rm 2}$ induced by air exposure}.
	\newblock \href {\doibase 10.1088/2053-1583/aab0cd} {
		\emph{\bibinfo{journal}{2D Materials}} \textbf{\bibinfo{volume}{5}},
		\bibinfo{pages}{025017} (\bibinfo{year}{2018})}.
	
	\bibitem{Houssa2019}
	\bibinfo{author}{Houssa, M.} \emph{et~al.}
	\newblock \bibinfo{title}{Contact resistance at graphene/{MoS}$_{\rm 2}$
		lateral heterostructures}.
	\newblock \href {\doibase 10.1063/1.5083133} {
		\emph{\bibinfo{journal}{Appl. Phys. Lett.}} \textbf{\bibinfo{volume}{114}},
		\bibinfo{pages}{163101} (\bibinfo{year}{2019})}.
	
	\bibitem{Chen2017}
	\bibinfo{author}{Chen, W.}, \bibinfo{author}{Yang, Y.}, \bibinfo{author}{Zhang,
		Z.} \& \bibinfo{author}{Kaxiras, E.}
	\newblock \bibinfo{title}{Properties of in-plane graphene/{MoS}$_{\rm 2}$
		heterojunctions}.
	\newblock \href {\doibase 10.1088/2053-1583/aa8313} {
		\emph{\bibinfo{journal}{2D Materials}} \textbf{\bibinfo{volume}{4}},
		\bibinfo{pages}{045001} (\bibinfo{year}{2017})}.
	
	\bibitem{Pandey2016}
	\bibinfo{author}{Pandey, M.} \emph{et~al.}
	\newblock \bibinfo{title}{Defect-tolerant monolayer transition metal
		dichalcogenides}.
	\newblock \href {\doibase 10.1021/acs.nanolett.5b04513} {
		\emph{\bibinfo{journal}{Nano Lett.}} \textbf{\bibinfo{volume}{16}},
		\bibinfo{pages}{2234--2239} (\bibinfo{year}{2016})}.
	
	\bibitem{Wang2019}
	\bibinfo{author}{Wang, Y.} \emph{et~al.}
	\newblock \bibinfo{title}{Van der {W}aals contacts between three-dimensional
		metals and two-dimensional semiconductors}.
	\newblock \href {\doibase 10.1038/s41586-019-1052-3} {
		\emph{\bibinfo{journal}{Nature}} \textbf{\bibinfo{volume}{568}},
		\bibinfo{pages}{70--74} (\bibinfo{year}{2019})}.
	
	\bibitem{English2016}
	\bibinfo{author}{English, C.~D.}, \bibinfo{author}{Shine, G.},
	\bibinfo{author}{Dorgan, V.~E.}, \bibinfo{author}{Saraswat, K.~C.} \&
	\bibinfo{author}{Pop, E.}
	\newblock \bibinfo{title}{{Improved Contacts to {MoS}$_{\rm 2}$ Transistors by
			Ultra-High Vacuum Metal Deposition}}.
	\newblock \href {\doibase 10.1021/acs.nanolett.6b01309} {
		\emph{\bibinfo{journal}{Nano Lett.}} \textbf{\bibinfo{volume}{16}},
		\bibinfo{pages}{3824--3830} (\bibinfo{year}{2016})}.
	
	\bibitem{Das2013}
	\bibinfo{author}{Das, S.} \& \bibinfo{author}{Appenzeller, J.}
	\newblock \bibinfo{title}{Where does the current flow in two-dimensional
		layered systems?}
	\newblock \href {\doibase 10.1021/nl401831u} {
		\emph{\bibinfo{journal}{Nano Lett.}} \textbf{\bibinfo{volume}{13}},
		\bibinfo{pages}{3396--3402} (\bibinfo{year}{2013})}.
	
	\bibitem{Schulman2018}
	\bibinfo{author}{Schulman, D.~S.}, \bibinfo{author}{Arnold, A.~J.} \&
	\bibinfo{author}{Das, S.}
	\newblock \bibinfo{title}{Contact engineering for {2D} materials and devices}.
	\newblock \href {\doibase 10.1039/C7CS00828G} {
		\emph{\bibinfo{journal}{Chem. Soc. Rev.}} \textbf{\bibinfo{volume}{47}},
		\bibinfo{pages}{3037--3058} (\bibinfo{year}{2018})}.
	
	\bibitem{Kappera2014}
	\bibinfo{author}{Kappera, R.} \emph{et~al.}
	\newblock \bibinfo{title}{Phase-engineered low-resistance contacts for
		ultrathin {MoS}$_{\rm 2}$ transistors}.
	\newblock \href {\doibase 10.1038/nmat4080} {
		\emph{\bibinfo{journal}{Nature Mat.}} \textbf{\bibinfo{volume}{13}},
		\bibinfo{pages}{1128--1134} (\bibinfo{year}{2014})}.
	
	\bibitem{Sung2017}
	\bibinfo{author}{Sung, J.~H.} \emph{et~al.}
	\newblock \bibinfo{title}{Coplanar semiconductor-metal circuitry defined on
		few-layer {MoTe}$_{\rm 2}$ via polymorphic heteroepitaxy}.
	\newblock \href {\doibase 10.1038/nnano.2017.161} {
		\emph{\bibinfo{journal}{Nature Nanotech.}} \textbf{\bibinfo{volume}{12}},
		\bibinfo{pages}{1064--1070} (\bibinfo{year}{2017})}.
	
\end{thebibliography}
\end{document}